\documentclass[aps,pra,floatfix,twocolumn,showpacs,amsmath,amssymb,letterpaper]{revtex4-2}
\usepackage{graphicx}
\usepackage{subfigure}
\usepackage{array}
\usepackage{bigstrut}
\usepackage{longtable}
\usepackage{rotating}
\usepackage{diagbox}
\usepackage[mode=text]{siunitx}
\usepackage{bm}
\usepackage{float}
\usepackage{lipsum}
\usepackage{warpcol}
\usepackage{color}
\usepackage{amsthm,amsmath}
\usepackage{mathrsfs}
\usepackage{appendix}
\usepackage{multirow}
\usepackage{threeparttable}
\usepackage{soul}
\usepackage[colorlinks=true, linkcolor=blue, citecolor=blue, urlcolor=blue]{hyperref}
\usepackage{nameref}
\usepackage{csquotes}
\usepackage{tabularx}
\usepackage{booktabs} % ?? \toprule, \midrule ?
\usepackage{dcolumn}
\usepackage{multirow}
\usepackage{makecell}
\allowdisplaybreaks

\newcolumntype{d}[1]{D{.}{.}{#1}}

\begin{document}
	
	\title{Low-energy positron scattering from metastable helium}
	
	\author{Ning-Ning Gao$^{1, 2}$}
	\author{Hui-Li Han$^{1}$}
	\email{huilihan@wipm.ac.cn}
	\author{Ting-Yun Shi$^{1}$}
	\author{Li-Yan Tang$^{1}$}
	
	\affiliation{
		$^{1}$
		%Wuhan Institute of Physics and Mathematics,
		Innovation Academy for Precision Measurement Science and Technology,
		Chinese Academy of Sciences, Wuhan 430071, People's Republic of China
	}
	
	\affiliation{
		$^{2}$University of Chinese Academy of Sciences, Beijing 100049, People's Republic of China
	}

	\date{\today}
	
	\begin{abstract}
Low-energy positron scattering from singlet and triplet metastable He($1s2s$) is investigated using the $R$-matrix propagation method in hyperspherical coordinates. Elastic and positronium-formation cross sections are reported, and near-threshold resonance structures are analyzed in terms of eigenphase sums and the time-delay matrix. For the triplet target, the calculated cross sections are in agreement with available convergent-close-coupling results and display the expected threshold behavior. Beyond the known $S$-wave features, Feshbach resonance series in higher partial waves extending up to highly excited atomic thresholds are systematically identified. The time-delay matrix further uncovers hidden resonances that produce obvious structures in the positronium-formation cross sections.

	\end{abstract}
	\maketitle
	\section{Introduction}
Positron scattering from atoms has long been a central topic in
antimatter and few-body collision physics. Extensive theoretical and
experimental studies have addressed elastic scattering, electronic
excitation, ionization, positronium (Ps) formation, and annihilation
over a broad range of impact energies~\cite{wymerexcitation2026,Gao2025Jul,2025PopoviczMay,
2025atoms13060046,Mori2024,2024MISTRY2024147470,Li2023Dec,Mori2023Mar,
2023atoms11040065,McEachran2019,ratnavelurecommended2019,Yan2018Dec,
Kadyrov2016,2015Zecca17112015,Brawley2015,Bray2014,KARWASZ2005666,Surko2005}.
Among these processes, Ps formation is especially important because it
opens a rearrangement channel and can strongly alter the low-energy
scattering dynamics.

Helium is one of the standard benchmark targets for positron-atom
collision studies. For ground-state helium, elastic scattering,
excitation, Ps formation, and resonance phenomena have been studied
extensively~\cite{Li2023Dec}. Resonant structures are of particular interest
because they may indicate temporary trapping of the positron and thus
provide indirect evidence for positron-atom quasibound states~\cite{Harabati2014}. Mitroy
and co-workers~\cite{Bromley2012Aug} predicted resonances in the positron-helium system near
doubly excited helium thresholds and predicted a shape resonance just above the
$\mathrm{He}(2s^2\,^{1}S^e)$ threshold at 57.8485 eV~\cite{Machacek2012Dec}. Yan and Ho later
used the complex-coordinate-rotation method to study triply excited
autodissociating resonant states of the positron-helium system~\cite{Yan2018Dec}.
Nevertheless, unambiguous experimental evidence for positron attachment
to an atomic target remains lacking.

Compared with ground-state helium, positron scattering from metastable
helium has been studied less extensively. Hanssen \textit{et al.}~\cite{Hanssen2000}
calculated Ps formation above 50 eV by including the Ps($1s$) channel
in the close-coupling expansion and found larger scattering and
rearrangement cross sections for metastable helium than for
ground-state helium. At lower energies, Utamuratov \textit{et al.}~\cite{Utamuratov2010Oct} performed systematically convergent two-center close-coupling (CCC) calculations for positron scattering from He($1s2s\,{}^{3}S$), explicitly including both atomic and Ps channels in the close-coupling expansion. These calculations provide an important benchmark for this system. The CCC calculation employed a frozen-core configuration-interaction model and showed that the low-lying triplet states, including the metastable $1s2s\,{}^{3}S$ state, were well described, thereby supporting the approximation adopted in the present work. Wu \textit{et al.}~\cite{Wu2017} also investigated the low-energy scattering using the momentum-space coupled-channel optical (CCO) approach. Noticeable differences were found between the CCC and CCO cross sections in the low-energy region. These differences highlight the importance of a rigorous treatment of interchannel coupling, particularly between the atomic and Ps-formation channels.

The identification of resonances in multichannel scattering requires careful analysis. For an isolated resonance with a slowly varying background, the
eigenphase sum follows the Breit-Wigner form and its resonant part
increases by approximately $\pi$~\cite{Hazi1979}. In positron-atom
systems, however, atomic channels,
$e^{\scriptscriptstyle +}+\mathrm{He}^{\scriptscriptstyle*}$, and rearrangement channels,
$\mathrm{Ps}+\mathrm{He}^{\scriptscriptstyle +}$, can be strongly
coupled, and the background phase may mask the resonant variation. In
positron scattering from He$^{\scriptscriptstyle +}$, stable complex-rotation eigenvalues
with large widths produced only weak changes in the eigenphase sum, but
clear peaks appeared in the trace or in a single eigenvalue of the
time-delay matrix~\cite{Igarashi2004}. Similar time-delay signatures
were found for broad resonances of H$^{\scriptscriptstyle -}$ and Ps$^{\scriptscriptstyle -}$ within the
hyperspherical framework~\cite{Igarashi2016}.

In this work we study low-energy positron scattering from
$\mathrm{He}(1s2s\,{}^1S)$ and $\mathrm{He}(1s2s\,{}^3S)$ using the
hyperspherical $R$-matrix propagation method. Elastic and Ps-formation
cross sections are calculated, and near-threshold resonances are
analyzed with both eigenphase sums and the time-delay matrix. Feshbach resonance series in higher partial waves extending up to highly excited atomic thresholds are systematically
 identified. The time-delay matrix further uncovers hidden resonances
whose eigenphase sums increase by much
less than $\pi$. These resonances are more clearly revealed by peaks in
the time-delay eigenvalues and produce visible structures in the
Ps-formation cross sections.

The remainder of this paper is organized as follows. Section~\ref{sec:method}
describes the theoretical method and model potentials. Section~\ref{sec:discussion}
presents the Ps($n=1$)--He$^{\scriptscriptstyle +}(1s)$ scattering
lengths, the threshold behavior of the cross sections, and the
partial-wave resonances in the
$e^{\scriptscriptstyle +}+\mathrm{He}(1s2s\,{}^1S)$ and
$e^{\scriptscriptstyle +}+\mathrm{He}(1s2s\,{}^3S)$ systems. Section~\ref{sec:summary}
summarizes the main conclusions. Atomic units are used throughout unless
otherwise stated.

\section{Theoretical Method}   %  <-- PRA style heading
\label{sec:method}
%==================================================================
\subsection{Hypersphrical coordinate method}
\begin{figure}[htbp]
	\centering
	\subfigure{
		\includegraphics[scale=0.22]{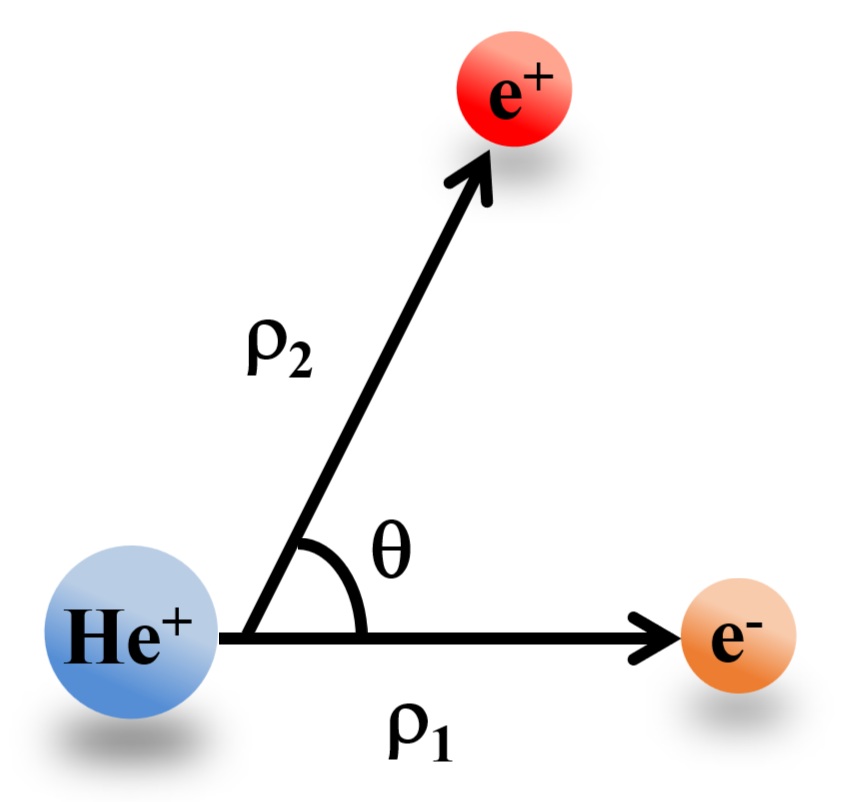}
	}
	\caption{The Jacobi coordinates for the $e^{\scriptscriptstyle+}$-He system.}
	\label{fig1}
\end{figure}

In this study, we investigate the collision properties of the $e^{\scriptscriptstyle+}$-He(1s2s) system. The metastable helium in the $\mathrm{He}(1s2s\,^1S)$ and $\mathrm{He}(1s2s\,^3S)$ states is represented as the single valence electron interacting with the core via a local central potential, and the positron-He(1s2s) atom system therefore reduces to the equivalent three-body system. The masses of the three charged particles, the He$^{\scriptscriptstyle+}$ core, $e^{\scriptscriptstyle-}$, and $e^{\scriptscriptstyle+}$, are denoted by $m_{1}$, $m_{2}$, and $m_{3}$, respectively. We employ Delves's hyperspherical coordinates and introduce the mass-scaled Jacobi coordinates~\cite{lin1995} . The first Jacobi vector $\vec{\rho}_{\scriptscriptstyle 1}$ is chosen to be the vector from the He$^{\scriptscriptstyle+}$ core to $e^{\scriptscriptstyle-}$, with reduced mass $\mu_{1}$, and the second Jacobi vector $\vec{\rho}_{2}$ goes from the diatom center of mass to $e^{\scriptscriptstyle+}$, with reduced mass $\mu_{2}$. The angle between $\vec{\rho}_{1}$ and $\vec{\rho}_{2}$ is denoted by $\theta$, as shown in the Figure~\ref{fig1}. The hyperradius $R$ and hyperangle $\phi$ are defined as\\
\begin{equation}
	\label{1}
	\mu R^{2}=\mu_{1}\rho_{1}^{2}+\mu_{2}\rho_{2}^{2}\,,
\end{equation}
and\\
\begin{equation}
	\label{2}
	\tan\phi=\sqrt{\frac{\mu_{2}}{\mu_{1}}}\frac{\rho_{2}}{\rho_{1}},\;\; 0 \leq\phi\leq\frac{\pi}{2}\,,
\end{equation}
respectively, where $R$ is the only coordinate with the dimension of length and represents the overall size of the three-body system. The rotation of the plane containing the three particles is described collectively by $\Omega$ $[\Omega \equiv (\theta, \phi, \alpha, \beta, \gamma)]$, which includes $\theta$, $\phi$, and three Euler angles $(\alpha, \beta, \gamma)$. The parameter $\mu$ is an arbitrary scaling factor, and we choose $\mu=\sqrt{\mu_{1}\mu_{2}}$ for our calculations.

The Schr$\mathrm{\ddot{o}}$dinger equation in hyperspherical coordinates can be written after rescaling the three-body wave function $\Psi_{\upsilon'}$ as $\psi_{\upsilon'}(R;\Omega)=\Psi_{\upsilon'}(R;\Omega) R^{5/2} \sin\phi \cos\phi$:\\
\begin{equation}
	\begin{aligned}
		\label{3}
		&\left[ -\frac{1}{2\mu}\frac{d^{2}}{dR^{2}}+\left( \frac{\Lambda^{2}-\frac{1}{4}}{2\mu R^{2}}+V(R;\theta,\phi)\right) \right] \psi_{\upsilon'}(R;\Omega)\\
		&=E\psi_{\upsilon'}(R;\Omega)\,,
	\end{aligned}
\end{equation}
where $\Lambda^{2}$ is the square ``grand angular momentum operator", and its expression is as given in Ref.~\cite{lin1995}. The three-body interaction potential $V(R;\theta,\phi)$ is expressed as:\\
\begin{equation}
	\label{4}
	V(R;\theta,\phi)=v_{e}(r_{12})+v_{p}(r_{ 13})+v_{ep}(r_{23})\,,
\end{equation}
where $r_{12}$, $r_{13}$, and $r_{23}$ are the electron-core distance, the positron-core distance, and the electron-positron distance, respectively.

The three-body wave function $\psi_{\upsilon'}$ can be expanded with the complete, orthonormal set of the angular wave function $\Phi_{\nu}$ and radial wave functions $F_{\nu\upsilon'}$ as
\begin{equation}
	\label{8}
	\psi_{\upsilon'}(R;\Omega)=\sum\limits_{\nu=0}^{\infty}F_{\nu\upsilon'}(R)\Phi_{\nu}(R;\Omega)\,.
\end{equation}
The adiabatic potentials $U_{\nu}(R)$ and channel functions $\Phi_{\nu}(R;\Omega)$ at fixed $R$ can be obtained by solving the following adiabatic eigenvalue equation:
\begin{equation}
	\label{9}
	\left( \frac{\Lambda^{2}-\frac{1}{4}}{2\mu R^{2}}+V(R;\theta,\phi)\right) \Phi_{ \nu}(R;\Omega)=U_{ \nu}(R) \Phi_{\nu}(R;\Omega)\,.
\end{equation}

The channel function is further expanded on Wigner rotation matrices $D_{KM}^{J}$ as
\begin{align}
	\label{10}
	\Phi_{\nu}^{J\Pi M}(R;\Omega)=\sum\limits_{K=0}^{J}u_{\nu K}(R;\theta,\phi)\overline{D}_{ KM}^{ J\Pi}(\alpha,\beta,\gamma)\,.
\end{align}
\begin{align}
	\label{11}
	\overline{D}_{KM}^{J\Pi}=\frac{1}{4\pi}\sqrt{2J+1}[D_{KM}^{J}+(-1)^{K+J}\Pi D_{-KM}^{J}]\,,
\end{align}
where $J$ is the total nuclear orbital angular momentum, $M$ is its projection onto the laboratory-fixed axis, and $\Pi$ is the parity with respect to the inversion of the nuclear coordinates. The quantum number $K$ denotes the projection of $J$ onto the body-frame $z$ axis and takes the values $K=J,J-2,\ldots,-(J-2),-J$ for the ``parity-favored" case, $\Pi=(-1)^{ J}$, and $K=J-1,J-3,\ldots,-(J-3),-(J-1)$ for the ``parity-unfavored" case, $\Pi=(-1)^{J+1}$. $u_{\nu K}(R;\theta,\phi)$ is expanded with $B$-spline basis sets:

\begin{align}
	\label{12}
	u_{\nu K}(R;\theta,\phi)=\sum\limits_{i}^{ N_{\phi}} \sum\limits_{j}^{N_{\theta}}c_{ i,j}B_{i}(\phi)B_{j}(\theta)\,,
\end{align}
where $N_{\theta}$ and $N_{\phi}$ are the sizes of the basis sets in the $\theta$ and $\phi$ directions, respectively. Using the $B$-splines as a basis function has multiple advantages, including high localization, flexibility, completeness, and numerical stability~\cite{Bachau2001,Kang2006}, which enable us to obtain accurate potential curves and channel functions by employing these advantages.

\subsection{$R-$matrix propagation}

The hyperradius $R$ is divided into $(N-1)$ intervals using a set of grid points $R_{ 1}<R_{ 2}<\cdots R_{ N}$. In the present calculations, 204 sectors are employed over the range $R=0.1\,a_0$ to $R=300\,a_0$. At short distances, we utilize the SVD method to solve Eq.~(\ref{3}) in the interval $[R_{ i},R_{i+1}]$. In this method, we express the total wave function $\psi_{\upsilon'}(R;\Omega)$ in terms of the discrete variable representation (DVR) basis $\pi_{i}$ and the channel functions $\Phi_{\nu}(R;\Omega)$ as follows:
\begin{align}
	\label{16}
	\psi_{\upsilon'}(R;\Omega)=\sum^{N_\text{DVR}}_{i}\sum^{ N_\text{chan}}_{\nu}C^{ \upsilon'}_{ i\nu}\pi_{ i}(R)\Phi_{\nu}(R_{i};\Omega)\,,
\end{align}
where $N_\text{DVR}$ represents the number of DVR basis functions and $N_\text{chan}$ is the number of included channel functions. The values $N_{\rm DVR}=37$ and $N_{\rm chan}=29$ are used. By inserting $\psi_{ \upsilon'}(R;\Omega)$ into the three-body Schr$\mathrm{\ddot{o}}$dinger equation given by Eq.\;(\ref{3}), we arrive at a standard algebraic problem for the coefficients $C^{\upsilon'}_{i\nu}$:
\begin{align}
	\label{17}
	\sum^{N_\text{DVR}}_{j}\sum^{N_\text{chan}}_{\mu}\mathcal{T}_{ij}
	\mathcal{O}_{i\nu,j\mu}C^{\upsilon'}_{ j\mu}+U_{\nu} (R_{i})C^{\upsilon'}_{i\nu}=E^{ \upsilon'}C^{ \upsilon'}_{i\nu}\,,
\end{align}
where
\begin{align}
	\label{18}
	\mathcal{T}_{ij}=\frac{1}{2\mu}\int^{R_{i+1}}_{ R_{ i}}\frac{d}{dR}\pi_{i}(R)\frac{d}{dR}\pi_{j}(R)dR\,,
\end{align}
are the kinetic-energy matrix elements, with ${R_{i}}$ and ${R_{i+1}}$ being the boundaries of the calculation box, and
\begin{align}
	\label{19}
	\mathcal{O}_{i\nu,j\mu}=\langle\Phi_{\nu}(R_{i};\Omega)|\Phi_{ \mu}(R_{j};\Omega)\rangle
\end{align}
are the overlap matrix elements between the adiabatic channels defined at different quadrature points.

At large distances, the traditional adiabatic hyperspherical method is used to solve Eq.~(\ref{3}).
When substituting the wave functions $\psi_{\upsilon'}(R;\Omega)$ into Eq.~(\ref{3}), one obtains a set of coupled ordinary differential equations:
\begin{equation}
	\begin{aligned}
		\label{20}
		&	\left[-\frac{1}{2\mu}\frac{d^{2}}{dR^{2}}+U_{\nu}(R)- E\right]F_{\nu,\upsilon'}(R)\\
		&-\frac{1}{2\mu}\sum_{\mu}\left[2P_{\mu\nu}(R)\frac{d}{dR}+Q_{\mu\nu}(R)\right]F_{\mu \upsilon'}(R)=0\,,
	\end{aligned}
\end{equation}
where
\begin{align}
	\label{21}
	P_{\mu\nu}(R)=\int d\Omega \Phi_{ \mu}(R;\Omega)^{\scriptscriptstyle *}\frac{\partial}{\partial R}\Phi_{\nu}(R;\Omega)\,,
\end{align}
and
\begin{align}
	\label{22}
	Q_{\mu\nu}(R) = \int d\Omega \Phi_{ \mu}(R;\Omega)^{\scriptscriptstyle *}\frac{\partial^{2}}{\partial R^{2}}\Phi_{\nu}(R;\Omega)\,.
\end{align}
are the nonadiabatic couplings that control the inelastic transitions as well as the width of the resonance supported by  adiabatic potential $U_{\nu}(R)$ .

The effective hyperradial potentials that include hyperradial kinetic energy contributions with the $P_{ \nu\nu}^{2}$ term are more physical than adiabatic hyperpotentials and are defined as
\begin{align}
	\label{23}
	W_{\nu \nu}(R)=U_{ \nu}(R)-\frac{\hbar^{2}}{2 \mu} P_{\nu \nu}^{2}(R)\,.
\end{align}

The effective potentials for the bound-state channels exhibit the following asymptotic behavior at large \( R \):
\begin{equation}
	\label{24}
	W_{\nu \nu}(R) = \frac{l_{\nu} (l_{\nu} +1)}{2\mu R^{2}} + E_{2b}\,,
\end{equation}
where \( E_{2b} \) is the two-body bound-state energy, and \( l_{\nu} \) represents the angular momentum of the third particle relative to the two-body bound system, whose angular momentum is denoted by \( l_{1} \).

In the hyperspherical method, the total angular momentum \( J \), along with \( l_{1} \) and \( l_{\nu} \), follows the selection rule:
\begin{equation}
	\label{25}
	|l_{1} - l_{\nu} | \leq J \leq l_{ 1} + l_{\nu}\,.
\end{equation}

The goal of our scattering study is to determine the scattering matrix $\underline{\mathcal{S}}$ from the solutions of Eq.~(\ref{3}). We first calculate the $\underline{\mathcal{R}}$ matrix, which is defined as
\begin{align}
	\label{13}
	\underline{\mathcal{R}}(R)=\underline{\textsl{F}}(R)[\widetilde{\underline{\textsl{F}}}(R)]^{ -1}\,,
\end{align}
where matrices $\underline{\textsl{F}}$ and $\widetilde{\underline{\textsl{F}}}$ can be calculated from the solution of Eqs.~(\ref{3}) and (\ref{9}) by evaluating the integrals:
\begin{align}
	\label{14}
	F_{\nu,\upsilon'}(R)=\int d\Omega \Phi_{\nu}(R;\Omega)^{\scriptscriptstyle *}\psi_{ \upsilon'}(R;\Omega)\,,
\end{align}
\begin{align}
	\label{15}
	\widetilde{F}_{\nu,\upsilon'}(R)=\int d\Omega \Phi_{\nu}(R;\Omega)^{\scriptscriptstyle *}\frac{\partial}{\partial R}\psi_{\upsilon'}(R;\Omega)\,.
\end{align}

We use the $R$-matrix propagation method. Within an interval $[R_{1},R_{2}]$, given an $\underline{\mathcal{R}}$ matrix at $R_{1}$, we calculate the corresponding $\underline{\mathcal{R}}$ matrix at another point $R = R_{2}$ using
\begin{align}
	\label{26}
	\underline{\mathcal{R}}(R_{2})=\underline{\mathcal{R}}_{22}
	-\underline{\mathcal{R}}_{21}\left[\underline{\mathcal{R}}_{11}
	+\underline{\mathcal{R}}(R_{1})\right]^{-1}\underline{\mathcal{R}}_{12}\,.
\end{align}
where the corresponding matrices give:
\begin{align}
	\label{27}
	(\underline{\mathcal{R}}_{11})_{\nu\mu} = \sum\limits_{n }\frac{u_{ \nu}^{(n)}(R_{1})u_{ \mu}^{(n)}(R_{1})}{2\mu (\varepsilon_{n}-E)}\,,
\end{align}
\begin{align}
	\label{28}
	(\underline{\mathcal{R}}_{12})_{\nu\mu} = \sum\limits_{n }\frac{u_{\nu}^{(n)}(R_{ 1})u_{\mu}^{(n)}(R_{2})}{2\mu (\varepsilon_{n}-E)}\,,
\end{align}
\begin{align}
	\label{29}
	(\underline{\mathcal{R}}_{21})_{\nu\mu} = \sum\limits_{n }\frac{u_{\nu}^{(n)}(R_{2})u_{ \mu}^{(n)}(R_{1})}{2\mu (\varepsilon_{n}-E)}\,,
\end{align}
\begin{align}
	\label{30}
	(\underline{\mathcal{R}}_{22})_{\nu\mu} = \sum\limits_{n }\frac{u_{\nu}^{(n)}(R_{2})u_{\mu}^{(n)}(R_{2})}{2\mu (\varepsilon_{n}-E)}\,,
\end{align}
where $\nu$ and $\mu$ denote different channels, indices 1 and 2 do not label the channel, and more details can be found in Ref.~\cite{WangJia2011}.

The $\underline{\mathcal{K}}$ matrix can be expressed as the following matrix equation:
\begin{align}
	\label{31}
	\underline{\mathcal{K}}=
	(\underline{f}-\underline{f}'\underline{\mathcal{R}})
	(\underline{g}-\underline{g}'\mathcal{R})^{-1}\,,
\end{align}
where $f_{\nu \nu'}=\sqrt{\frac{2\mu k_{\nu}}{\pi}} R j_{l_{\nu}}(k_{\nu} R)\delta_{\nu \nu'}$ and $g_{\nu \nu'}=\sqrt{\frac{2\mu k_{\nu}}{\pi}} R n_{l_{\nu}}(k_{ \nu} R)\delta_{\nu \nu'}$ are the diagonal matrices of energy-normalized spherical Bessel and Neumann functions.
For the bound-state channel, $l_{\nu}$ is the angular momentum of the third atom relative to the dimer and $k_{\nu}$ is given by $k_{\nu}=\sqrt{2 \mu\left(E-E_{2 b}\right)}$.
For the continuous channel, $l_{\nu}=\lambda_{\nu}+3 / 2$, and $k_{\nu}=\sqrt{2 \mu E}$. The scattering matrix $\underline{\mathcal{S}}$ is related to $\underline{\mathcal{K}}$ as follows:
\begin{align}
	\label{32}
	\underline{\mathcal{S}}=(\underline{1}+i\underline{\mathcal{K}})(\underline{1}-i\underline{\mathcal{K}})^{-1}\,.
\end{align}

The cross-sections are expressed in terms of the $\underline{\mathcal{S}}$ matrix as:
\begin{align}
	\label{33}
	\sigma_{ij}=\sum\limits_{J,\Pi}\frac{(2J+1)\pi}{k_{ad}^{2}}|S_{ f\leftarrow i}^{J,\Pi}-\delta_{ij}|^{2}\,.
\end{align}
where $i$ denotes the incident channel (the initial state) and $j$ denotes the exit channel (the final state). $k_{ad}=\sqrt{2\mu_{ad}(E-E_{2b})}$ is the atom-dimer wave number, and $\mu_{ad}$ is the atom-dimer reduced mass.

\subsection{Two-body interaction potential}

The model potential used for the valence electron and the core is from Ref.~\cite{Ryzhikh1998} and has the form

\begin{equation}
	\label{ve3S}
	v_{e}(r_{12})=v_{\text{dir}}(r_{12})+v_{\text{exc}}(r_{12})+v_{\text{pol}}(r_{12})\,.
\end{equation}
Unlike the model potential used in Ref.~\cite{Ren2011}, the present form separates the direct and exchange contributions, thereby excluding any exchange term from the interaction between the positron and the He$^{\scriptscriptstyle+}(1s)$ ionic core.

The direct interaction $v_{\text{dir}}(r)$ between the core and the active electron is computed from the exact He$^{\scriptscriptstyle+}(1s)$ wavefunction, which can be expressed in
\begin{equation}
	\label{ve}
	v_{\text{dir}}(r)=e^{-4r}\left(-\frac{1}{r}-2\right)-\frac{1}{r}\,.
\end{equation}

The polarization potential $v_{\text{pol}}(r)$ is
\begin{equation}
	\label{vpol}
	v_{\text{pol}}(r)=-\frac{\alpha_d}{2r^4}g^2(r)\,.
\end{equation}
where $g(r)=1-\exp\left(-\frac{r^6}{\rho^6}\right)$ is a cut-off function, which prevents the potential from diverging at the origin. The static dipole polarizability of He$^{\scriptscriptstyle+}(1s)$ is $\alpha_d=0.28125\, a_{0}^3$, the cut-off parameter of $\rho=2.0\, a_0$ was set as in Ref.~\cite{Mitroy2005} .

The core-exchange potential $v_{\text{exc}}(r)$ is determined empirically
\begin{equation}
	\label{vexc}
	v_{\text{exc}}(r)=(ar+br^2+c)e^{-\beta_1r}\,.
\end{equation}
Here, $v_{\rm exc}(r)$ is a state-dependent local effective potential
introduced to represent the short-range exchange effect between the
active electron and the frozen $1s$ core. For the
$e^{\scriptscriptstyle +}$--He($1s2s\,^{3}S$) system, the parameters are
taken from Ref.~\cite{Ren2012mar} as
$a=-0.96$, $b=-0.0167$, $c=-2.044$, and $\beta_1=2.8$.
For the $e^{\scriptscriptstyle +}$--He($1s2s\,^{1}S$) system, the
parameters $a$, $b$, $c$, and $\beta_1$ are adjusted to reproduce the
binding energies of the He atom, yielding
$a=0.96$, $b=0.0167$, $c=2.18$, and $\beta_1=5.5$.
The opposite signs of the singlet and triplet parameters originate from
the different exchange symmetries of the corresponding two-electron
wavefunctions. Consequently, the effective short-range exchange
correction is repulsive (positive) for the singlet state and attractive
(negative) for the triplet state.

The model potentials were first tested by calculating the ground- and excited-state energies of He. The Hamiltonian was diagonalized in a B-spline basis, and the resulting energies are listed in Table~\ref{t1}, together with the corresponding experimental values. As shown in the table, the present model potentials reproduce the experimental energies with reasonable accuracy. To further assess the quality of the corresponding wave functions, we also calculated oscillator strengths for several representative singlet and triplet transitions. The calculated triplet oscillator strengths agree with the reference values to within about 0.94\%, 0.96\%, and 0.089\% for the $1s2s\,^{3}S \to 1s2p\,^{3}P$, $1s2p\,^{3}P \to 1s3d\,^{3}D$, and $1s3d\,^{3}D \to 1s4f\,^{3}F$ transitions, respectively. For the corresponding singlet transitions, the deviations are about 18.21\%, 4.06\%, and 0.023\%, respectively. These comparisons provide an additional assessment of the quality of the model wave functions beyond their ability to reproduce the bound-state energies.

The positron–core interaction has no core-exchange potential. The core-direct interaction has the opposite sign, while the one-body polarization potential is the same. The positron-core interaction $v_{p}(r_{13})$ can be expressed as

\begin{equation}
	\label{vp3S}
	v_{p}(r_{13})=-v_{\text{dir}}(r_{13})+v_{\text{pol}}(r_{13})\,.
\end{equation}

The two-body potential between the positron and valence electron is
\begin{equation}
	\label{vep}
	v_{ep}(r_{23})=-\frac{1}{r_{23}}+\frac{\alpha_d}{r_{12}^2r_{13}^2}\cos\theta g(r_{12})g(r_{13})\,.
\end{equation}

%-------------------------------------------------------------------

{\renewcommand{\arraystretch}{1.35}
	\begin{table}
		\small
		\caption{\label{t1}
			Energies (a.u.) for helium singlet states calculated using a model potential and compared with available experimental data~\cite{NISTASD2024}, together with length-form oscillator strengths for selected He transitions. All energies are given relative to the He$^{\scriptscriptstyle+}(1s)$ ground state.}
		\begin{ruledtabular}
			\begin{tabular}{ccc}
				
				\multicolumn{3}{c}{State energies}\\
				\multicolumn{1}{c}{State}
				&\multicolumn{1}{c}{This work}
				&\multicolumn{1}{c}{Experiment~\cite{NISTASD2024}}\\
				\hline
				
				$1s2s$ & $-0.145367$ & $-0.145954$\\
				$1s3s$ & $-0.061254$ & $-0.061264$\\
				$1s4s$ & $-0.033596$ & $-0.033582$\\
				$1s2p$ & $-0.126641$ & $-0.123822$\\
				$1s3p$ & $-0.056032$ & $-0.055138$\\
				$1s4p$ & $-0.031451$ & $-0.031065$\\
				$1s3d$ & $-0.055623$ & $-0.055613$\\
				
				\hline
				\multicolumn{3}{c}{Length-form oscillator strengths}\\
				\multicolumn{1}{c}{Transition}
				&\multicolumn{1}{c}{This work}
				&\multicolumn{1}{c}{Ref.~\cite{Drake2023}}\\
				\hline				
				$1s2s\,^3S \to 1s2p\,^3P$
				& $0.5441639$ & $0.5390861$\\
				$1s2p\,^3P \to 1s3d\,^3D$
				&$ 0.6160550$&$0.6102252$\\
				$1s3d\,^3D \to 1s4f\,^3F$
				& $1.0134374$ & $1.0143389$\\
				$1s2s\,^1S \to 1s2p\,^1P$
				& $0.3078975$ & $0.3764403$\\
				$1s2p\,^1P \to 1s3d\,^1D$
				& $0.6813588$ & $0.7101641$\\
				$1s3d\,^1D \to 1s4f\,^1F$
                & $1.0153179$ & $1.0150829$\\				
			\end{tabular}
		\end{ruledtabular}
	\end{table}
}

\section{Results and Discussions}
\label{sec:discussion}

\subsection{Low-energy scattering parameters in the Ps($1s$)-He$^{\scriptscriptstyle+}$($1s$) channel}
\begin{figure*}[htbp]
	\centering
	\subfigure{
		\includegraphics[scale=0.3]{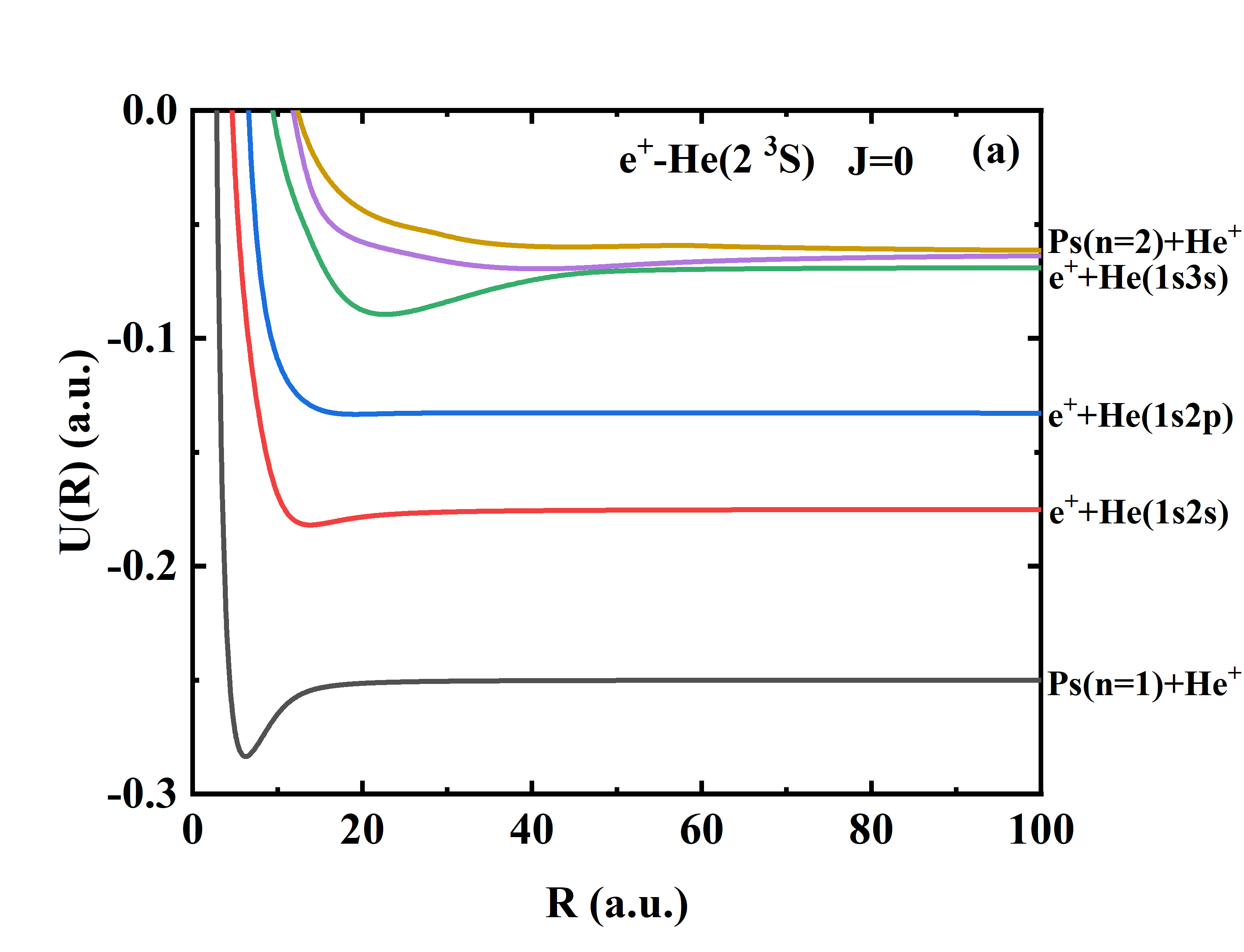}
		\label{fig2a}
	}
	\subfigure{
		\includegraphics[scale=0.3]{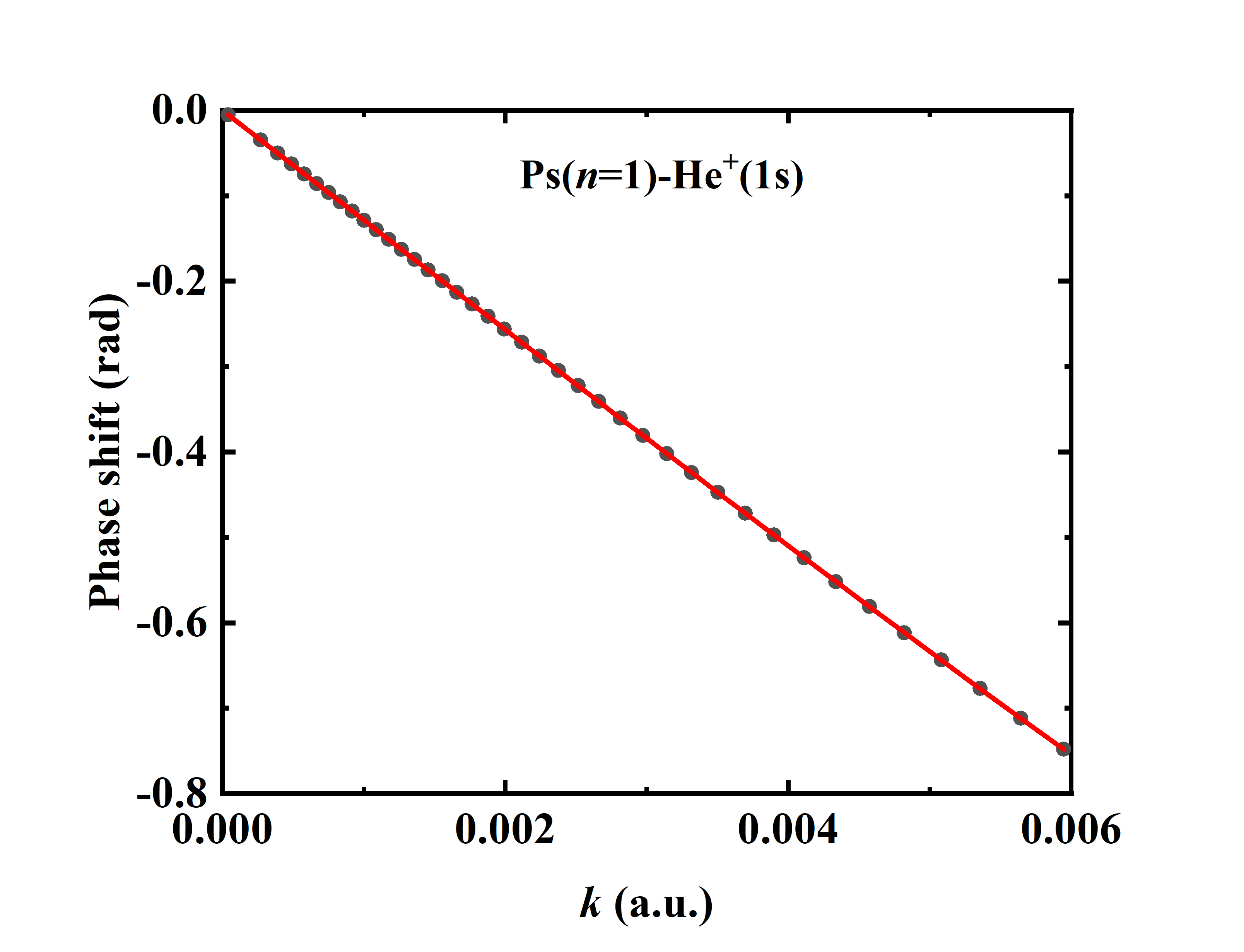}
		\label{fig2b}
	}
	\caption{(Color online) (a) The lowest several adiabatic potential curves for $J=0$ in the
		$e^{\scriptscriptstyle+}+\mathrm{He}(1s2s\,^3S)$ system. (b) $s$-wave phase shifts for Ps($1s$)-He$^{\scriptscriptstyle+}$($1s$) in the $e^{\scriptscriptstyle+}$-He($1s2s\,^{3}S$) system. The solid line is fitted by the formula~\eqref{eq:MERT}. }
	\label{fig2}
\end{figure*}

Figure~\ref{fig2a} shows the lowest adiabatic potential curves for $J=0$ in the $e^{\scriptscriptstyle +}+\mathrm{He}(1s2s\,^3S)$ system. The results were obtained with $N_{\theta}=76$ and $N_{\phi}=218$, for which the lowest 29 adiabatic curves are converged to at least six significant digits.
The Ps($1s$)--He$^{\scriptscriptstyle+}(1s)$ scattering length can be extracted
from the lowest potential curve correlated with the
Ps($1s$)+He$^{\scriptscriptstyle+}(1s)$ threshold. This quantity provides an
important low-energy parameter for the rearrangement channel.
To determine it, we calculated the low-energy phase shifts and
fitted them using the modified effective-range theory
(MERT)~\cite{1963PhysRev.130.1020}.

\begin{equation}
		\label{eq:MERT}
	\begin{split}
k\cot\delta(k) &=
-\frac{1}{A}
+\frac{\alpha_\text{eff} \pi k}{3A^2}
+\frac{2\alpha_\text{eff}k^2}{3A}\ln \left(\frac{\alpha_\text{eff} k^2}{16}\right)\\
&+Bk^2
+Ck^3+\mathcal{O}(k^4),
\end{split}
\end{equation}
where $k$ is the Ps($1s$)-He$^{\scriptscriptstyle+}$ relative momentum, $A$ is the scattering length, and $\alpha_\text{eff}=\mu\alpha_d = 72\,a_0^3$ is the effective dipole polarizability of Ps($1s$)~\cite{Mitroy2001}. The computed s-wave
phase shifts and the MERT fits are shown in Fig.~\ref{fig2b}. Using phase shifts in the range $k=4.0\times 10^{-5}$--0.006 $a_{0}^{-1}$, the extracted Ps($1s$)-He$^{\scriptscriptstyle+}(1s)$ scattering length is
$A=128\,a_0$ for the triplet metastable-helium system. The long-range interaction between Ps($1s$) and He$^{\scriptscriptstyle+}(1s)$ is governed by charge-induced dipole potential~\cite{Gao2013Aug},
\begin{equation}
	V(R)=-\frac{\alpha_d}{2r^4}\equiv-\frac{C_4}{r^4}\,,
\end{equation}	
where $C_4=\alpha_d/2$. The corresponding length scale is defined as
\begin{equation}
\beta_4=(2\mu C_4/\hbar^2)^{1/2}\,.
\end{equation}
with $\mu$ denotes the reduced mass of the Ps($1s$)-He$^{\scriptscriptstyle+}(1s)$ system. For the present system, $\beta_4 \approx 8.5\,a_0$. The Ps($1s$)--He$^{\scriptscriptstyle +}(1s)$ scattering length is therefore large on the polarization length scale, reflecting a strong near-threshold interaction in the rearrangement channel.
 This behavior is consistent with the recent
calculations of Zhao \textit{et al.}~\cite{cg41-dny5}, who predicted
positron attachment to the metastable triplet state
$\mathrm{He}(1s2s\,{}^3S)$. Their results show that the attached state is
highly diffuse and has a dominant
Ps($1s$)+He$^{\scriptscriptstyle +}(1s)$ character. In the
hyperspherical representation, this corresponds to a state mainly
supported by the adiabatic potential curve correlated with this
rearrangement threshold.

\subsection{Near-threshold elastic scattering and Ps formation in metastable helium}

We now turn to the elastic and inelastic cross sections for positron scattering from metastable $\mathrm{He}(1s2s\,^3S)$ and compare them with the available theoretical results. Previous calculations show noticeable differences in the low-energy region. For the triplet metastable target, the main previous low-energy studies are those of Utamuratov~\textit{et al.}~\cite{Utamuratov2010Oct}, based on the two-center CCC method, and those of Wu~\textit{et al.}~\cite{Wu2017}, obtained within the CCO approach.

\begin{figure*}[htbp]
	\centering
	\subfigure{
		\includegraphics[scale=0.3]{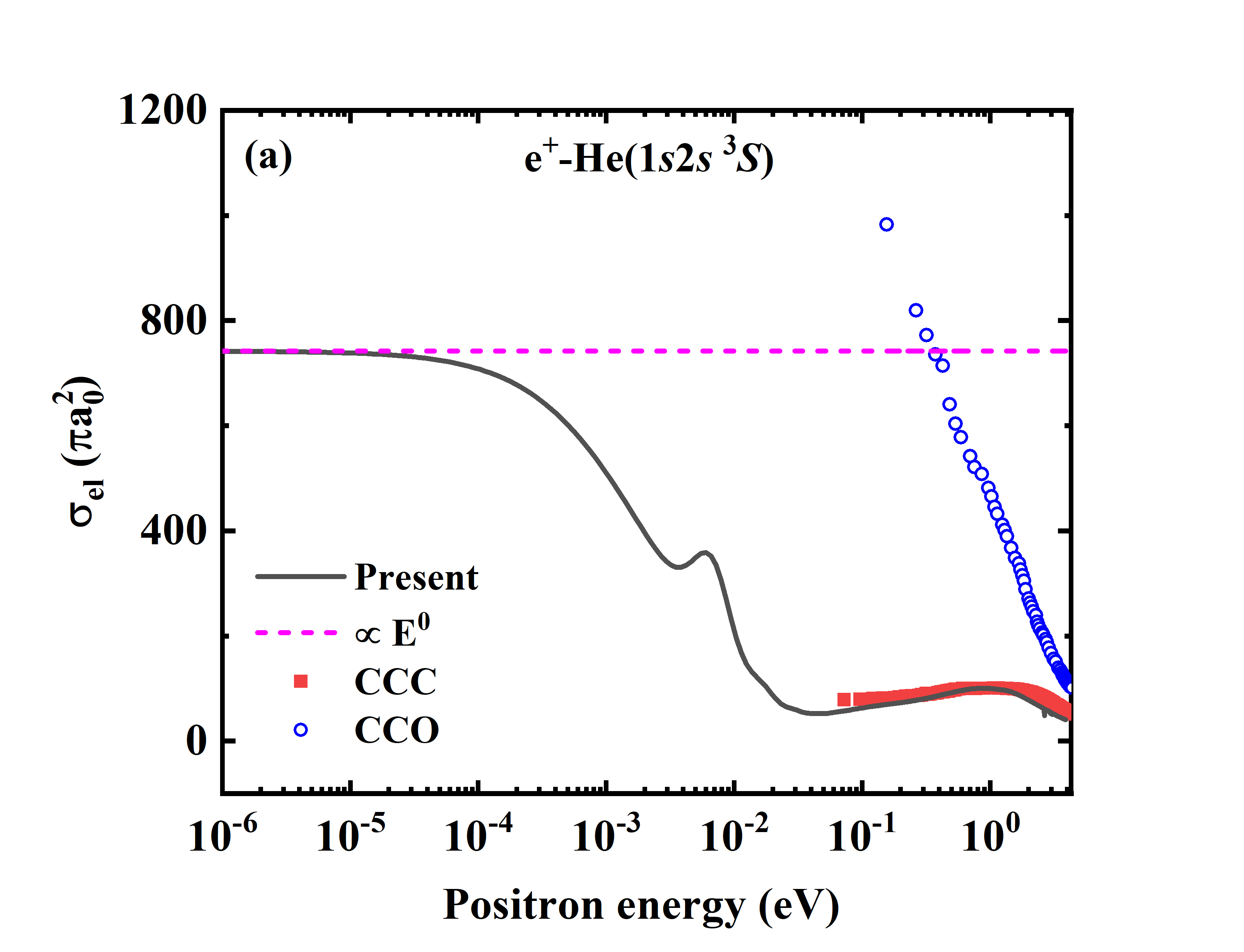}
		\label{fig3a}
	}
	\subfigure{
		\includegraphics[scale=0.3]{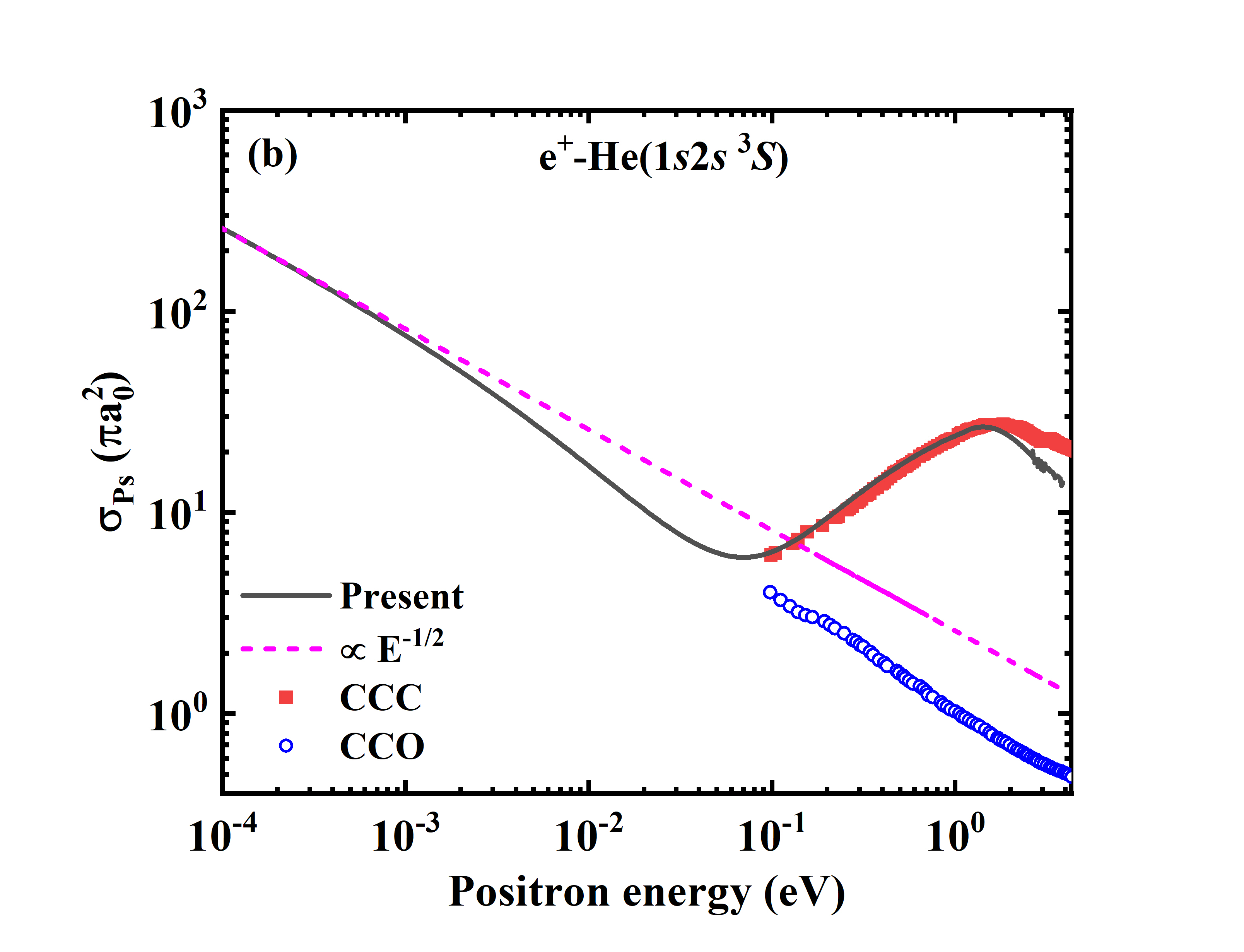}
		\label{fig3b}
	}
	\caption{(Color online) (a) Elastic scattering cross sections and (b) total Ps-formation cross sections for $e^{\scriptscriptstyle+}$-He($1s2s\,^3S$) collisions. The present results (solid black lines) are compared with the CCC~\cite{Utamuratov2010Oct} (red squares) and CCO~\cite{Wu2017} (open blue circles) results. The magenta dashed lines show the expected threshold laws, namely $\sigma_{\rm el}\propto E^{0}$ and $\sigma_{\rm Ps}\propto E^{-1/2}$, at ultralow collision energies.}
	\label{fig3}
\end{figure*}

Figure~\ref{fig3}(a) compares the present elastic cross section with
available theoretical results for positron scattering from
$\mathrm{He}(1s2s\,{}^3S)$. The present calculation is in good agreement
with the CCC results over the low-energy region shown, whereas the CCO
approach gives substantially larger elastic cross sections near
threshold. Figure~\ref{fig3}(b) shows the corresponding comparison for
Ps formation.
The present Ps-formation cross sections show good overall
agreement with the CCC results over their common energy range and
continue smoothly into the lower-energy region.

The expected threshold behavior, represented by the magenta dashed line in Fig.~\ref{fig3}, is well reproduced in the present calculations: the elastic cross section $\sigma_{\rm el}$ approaches a constant as
$E\to 0$, whereas the exothermic Ps-formation cross section follows
$\sigma_{\rm Ps}\propto E^{-1/2}$. The latter behavior follows directly
from the $S$-matrix expression for a rearrangement transition. For
Ps formation, $f\ne i$, and
\begin{equation}
    \sigma^{J\Pi}_{f\leftarrow i}
    =
    \sum\limits_{J}\frac{(2J+1)\pi}{k_i^2}
    \left|S^{J\Pi}_{f\leftarrow i}\right|^2 ,
\end{equation}
where $k_i$ is the incident wave number. The Wigner threshold law~\cite{wigner1948behavior} for an exothermic
inelastic or rearrangement process gives
$\left|S^{J\Pi}_{f\leftarrow i}\right|^2\propto k_i^{2l_i+1}$, where
$l_i$ is the orbital angular momentum in the incident channel. Hence
$\sigma^{J\Pi}_{f\leftarrow i}\propto k_i^{2l_i-1}$. The leading
contribution at threshold is the incident $s$ wave, giving
$\sigma_{\rm Ps}\propto k_i^{-1}\propto E^{-1/2}$. This is consistent
with the numerical results of Ref.~\cite{Gao2025Jul}, where the same
$E^{-1/2}$ threshold dependence was found for exothermic Ps formation in
low-energy positron-atom scattering.

\begin{figure*}[htbp]
	\centering
	\subfigure{
		\includegraphics[scale=0.3]{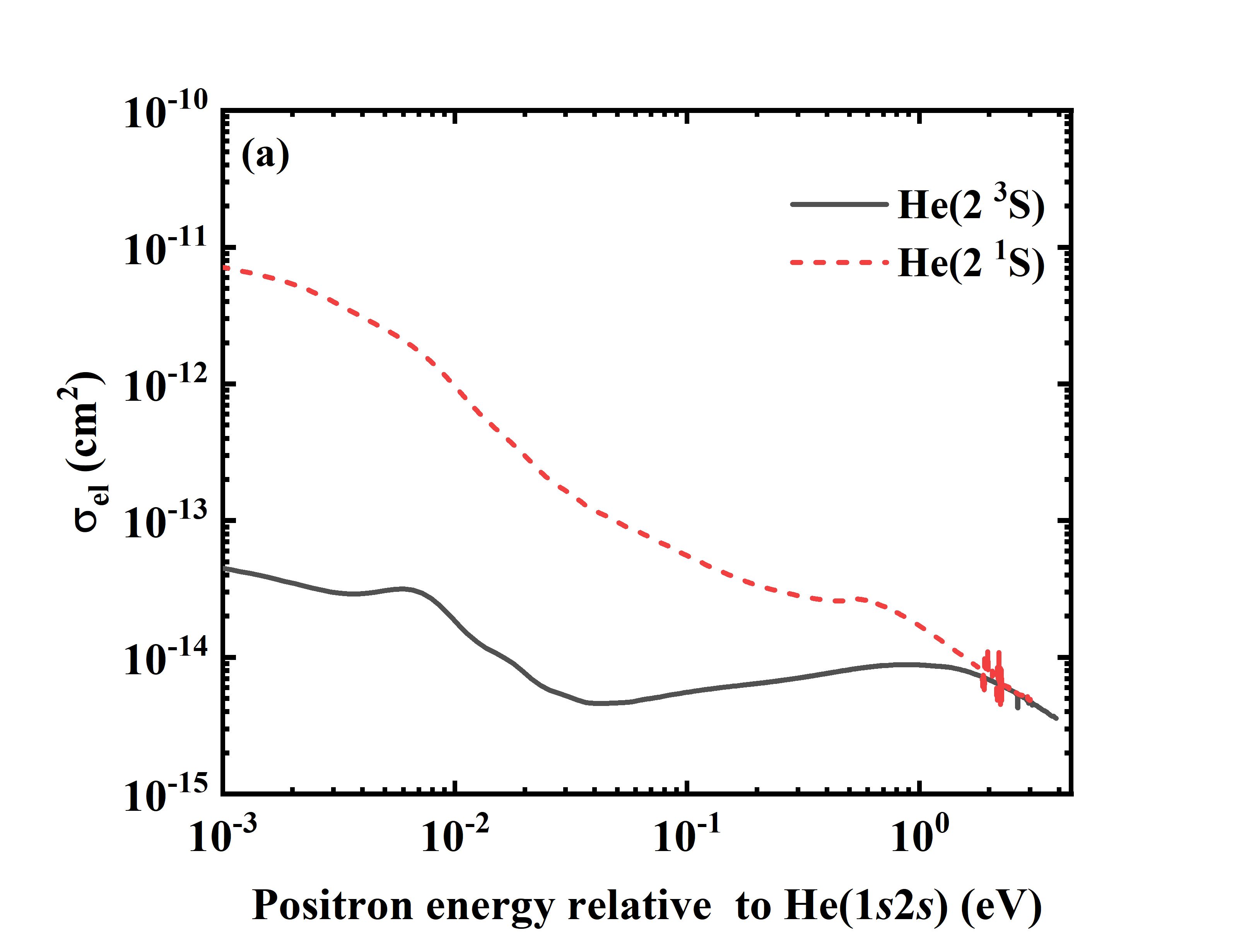}
		\label{fig4a}
	}
	\subfigure{
		\includegraphics[scale=0.3]{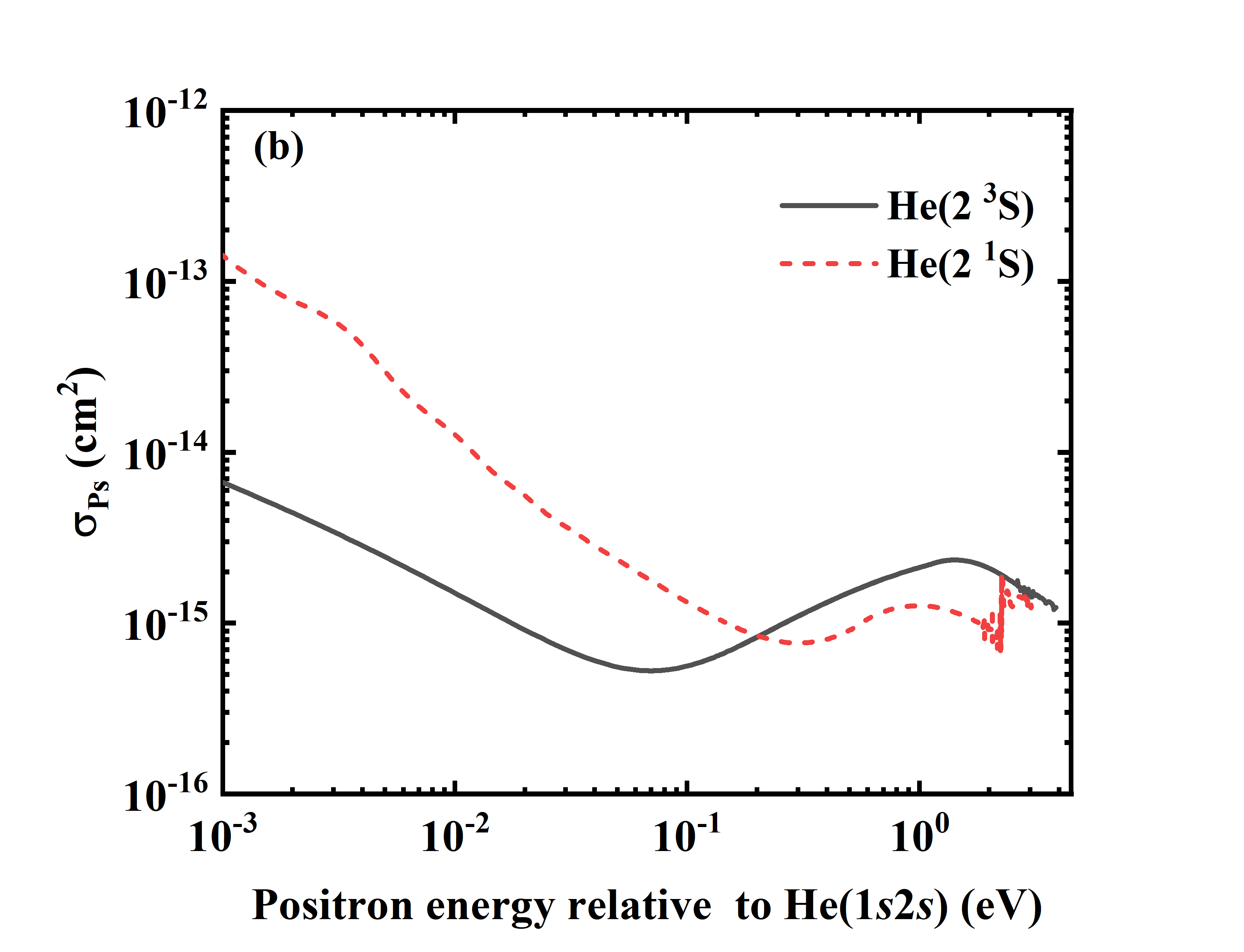}
		\label{fig4b}
	}
	\caption{(Color online) Comparison of (a) elastic and (b) Ps formation cross sections for positron scattering from metastable He($1s2s\,^1 S$) and He($1s2s\,^3 S$).}
	\label{fig4}
\end{figure*}

Figures~\ref{fig4a} and \ref{fig4b} show the elastic and Ps-formation
cross sections for positron scattering from
$\mathrm{He}(1s2s\,{}^1S)$ and $\mathrm{He}(1s2s\,{}^3S)$. In both
processes, the singlet cross sections are substantially larger than the
triplet ones over most of the low-energy region, especially at ultralow
collision energies. This trend is consistent with the much
larger static dipole polarizability of $\mathrm{He}(1s2s\,{}^1S)$. The
reported values are $\alpha_d=800.31633$ a.u. for
$\mathrm{He}(1s2s\,{}^1S)$ and $\alpha_d=315.6315(2)$ a.u. for
$\mathrm{He}(1s2s\,{}^3S)$~\cite{Yan052502Oct,Zhang2015Jul}. At low
energies, the positron--atom interaction is strongly influenced by the
long-range polarization potential. The larger polarizability of the singlet state therefore gives rise to a
stronger attractive interaction in the entrance channel.

\subsection{Resonances in positron scattering from metastable helium}

\subsubsection{Eigenphase-sum and time-delay analyses}

To determine the resonance energies $E_r$ and widths $\Gamma$, we use
the eigenphase-sum method as the primary diagnostic for well-isolated
resonances. The eigenphase shifts $\delta_i(E)$ are obtained by
diagonalizing the $\mathcal{K}$ matrix in Eq.~(\ref{31}) and taking the
arctangent of its eigenvalues. The total eigenphase sum is then given
by~\cite{Gao1989,Greene1987},
which is defined as
\begin{equation}
	\begin{aligned}
\delta_{\rm tot}(E)&=\sum_{i=1}^{N_o}\tan^{-1}\lambda_i\,,\\ &=\sum\limits_{i=1}^{N_o}\tan^{-1}\left[\frac{\Gamma_{i}}{2(E_{r}-E)}\right]+\sum\limits_{i=0}^{N_o}a_i(E)^i\,.
	\end{aligned}
\end{equation}
where $\lambda_i$ is the $i$th eigenvalue of the $\mathcal{K}$ matrix,
$E$ is the collision energy, and $N_o$ is the number of open channels. The term $\sum\limits_{i=0}^{N_o}a_i(E)^i$ represents the background contribution to the eigenphase sum. For an isolated resonance with a slowly varying background, the
eigenphase sum follows the Breit-Wigner behavior and increases by nearly
$\pi$ radians across the resonance region. In this case, $E_r$ and
$\Gamma$ can be extracted reliably from the energy dependence of
$\delta_{\rm tot}(E)$.
When the background eigenphase varies rapidly, the resonant increase of
$\delta_{\rm tot}(E)$ may be partly cancelled by the background
variation, so that the eigenphase sum does not display a full
$\pi$-radian rise. We therefore also analyze the time-delay matrix,
which provides an independent signature of resonance formation. The
time-delay matrix is defined by Smith as~\cite{1960PhysRev.118.349}
\begin{equation}
\label{Q}
Q(E)=i\hbar S(E)\frac{dS^{\dagger}(E)}{dE},
\end{equation}
where $S(E)$ is the open-channel scattering matrix and the dagger denotes
Hermitian conjugation. Its trace is related to the eigenphase sum and to
the eigenvalues $q_i(E)$ of $Q$ by
\begin{equation}
\label{TrQ}
{\rm Tr}\,Q(E)
=
2\hbar \frac{d\delta_{\rm tot}(E)}{dE}
=
\sum_i q_i(E).
\end{equation}
In the single-channel limit, $S(E)=\exp[2i\delta(E)]$, and
Eq.~(\ref{Q}) reduces to the Wigner time delay,
$Q(E)=2\hbar d\delta(E)/dE$. In a multichannel problem, the diagonal
element $Q_{ii}$ gives the time delay averaged over all final channels
for an incoming wave in channel $i$.

When a resonance is not clearly resolved by the eigenphase sum, either
${\rm Tr}\,Q(E)$ or one of the time-delay eigenvalues $q_i(E)$ may still
show a Lorentzian-like peak near $E=E_r$. The resonance parameters are
then obtained by fitting the time-delay signal to
\begin{equation}
\label{fittingTrQ}
{\rm Tr}\,Q(E)
\simeq
\frac{\hbar\Gamma}
{(E-E_r)^2+(\Gamma/2)^2}
+ C \,,
\end{equation}
where $C$ represents a background contribution. At the
resonance position, the resonant term has the approximate height ${\rm Tr}\,Q(E_r) \simeq \frac{4\hbar}{\Gamma}$.
Thus narrow isolated resonances produce sharp and high time-delay peaks,
whereas resonances embedded in a rapidly varying background may show
only a weak net change in $\delta_{\rm tot}(E)$. In the latter case,
peaks in ${\rm Tr}\,Q(E)$ or in individual time-delay eigenvalues provide
a more reliable identification of the resonance.

\subsubsection{Narrow Feshbach resonances}

There are two classes of asymptotic channels in the present systems.
The first consists of atomic channels, which dissociate into a positron
and an excited helium atom,
$e^{\scriptscriptstyle +}+\mathrm{He}^{\scriptscriptstyle *}$. The
second consists of rearrangement channels, which dissociate into a Ps
atom and a $\mathrm{He}^{\scriptscriptstyle +}$ ion. In the present
calculations, the atomic-channel expansion includes excited-helium
thresholds up to
$\mathrm{He}(1s4p)+e^{\scriptscriptstyle +}$, while the rearrangement
expansion includes all
$\mathrm{He}^{\scriptscriptstyle +}+\mathrm{Ps}(n\leq 2)$ channels.

We first identify a set of conventional narrow resonances. These states
show the standard Breit-Wigner behavior: the eigenphase sum increases by
nearly $\pi$ radians across each resonance region. Their resonance
energies and widths can therefore be extracted reliably from the
eigenphase-sum analysis. The resulting positions and total widths for
the $e^{\scriptscriptstyle +}$--$\mathrm{He}(1s2s\,{}^3S)$ and
$e^{\scriptscriptstyle +}$--$\mathrm{He}(1s2s\,{}^1S)$ systems are
listed in Table~\ref{t2}. The table also includes available results obtained by other methods~\cite{Ren2012mar,Ren2011} for comparison.
In addition, the present calculations predict a number of previously unreported resonance positions and higher partial-wave resonances.

For the $e^{\scriptscriptstyle +}$--He($1s2s\,{}^{1}S$) system,
S-wave resonances have been studied using the stabilization method.
Ren \textit{et al.}~\cite{Ren2011} investigated these resonances
within the hyperspherical coordinate framework, while Strasburger
\cite{Strasburger2018} further studied them using an all-electron
stabilization calculation with explicitly correlated Gaussian
functions. 
The present interaction model differs from that of
Ref.~\cite{Ren2011} by excluding the positron--nucleus exchange term,
providing a more consistent description of the positron--He$^{\scriptscriptstyle+}$
interaction. As shown in Table~\ref{t2}, the present resonance
parameters are in reasonable agreement with both previous stabilization
calculations.

For the $e^{\scriptscriptstyle +}$--He($1s2s\,{}^{3}S$) system,
the same interaction model as in Ref.~\cite{Ren2012mar} is adopted.
The $S$-wave resonance parameters available for this system were
reported by Ren \textit{et al.}~\cite{Ren2012mar} using the
stabilization method. The present results are in good overall
agreement with these resonance positions and widths.

Below the Ps($n=2$) thresholds, the resonances form Feshbach series induced by the long-range dipole interaction in the rearrangement channel. As shown in the Appendix, this interaction leads to an effective $-1/r^{2}$ asymptotic form, which supports a sequence of quasibound states converging to the Ps($n=2$) threshold.

The resonance positions satisfy an exponential scaling relation~\cite{Gailitis1963},
\begin{align}
	\label{Evratio}
	\frac{E_{\nu}}{E_{\nu+1}}=e^{\frac{2\pi}{\alpha}}\,.
\end{align}
where the slope parameter $\alpha$ depends on the partial wave and the subscript $\nu$ denotes different states in the series. The eigenphase-sum spectra and the corresponding structures in the positronium-formation cross sections are presented in the Appendix.

{\renewcommand{\arraystretch}{1.2}
	\begin{table*}[ht]
		\centering
		\caption{\label{t2} Resonance energies $E_{R}$ and widths $\Gamma$ of the lowest five broad partial-wave resonances in the $e^{\scriptscriptstyle+}$–He($1s2s\,^3S$) and $e^{\scriptscriptstyle+}$–He($1s2s\,^1S$) systems. Threshold energies are also listed. Results for the $S$-wave resonances are compared with those of Ren \textit{et al.}~\cite{Ren2012mar,Ren2011}. The notation x[y] means $x\times10^{-y}$. The values in blue bold denote the broad resonant structures of width of order $10^{-4}$ a.u., which produce pronounced structures in the Ps-formation cross sections.}
		\begin{ruledtabular}
			%\resizebox{0.5\textwidth}{!}{
			\begin{tabular}{cccccccccc}	
				%\hline\hline	
				\multicolumn{5}{c}{$e^{\scriptscriptstyle+}$–He($1s2s\,^{3}S$)}&\multicolumn{5}{c}{$e^{\scriptscriptstyle+}$–He($1s2s\,^{1}S$)}\\
				\cline{1-5} \cline{6-10}	
				&\multicolumn{2}{c}{Present}&\multicolumn{2}{c}{Ref.~\cite{Ren2012mar}}&&\multicolumn{2}{c}{Present}&\multicolumn{2}{c}{Literature}\\
				\cline{2-3} \cline{4-5} \cline{7-8}	\cline{9-10}	
				\multicolumn{1}{c}{Partial wave}&
				\multicolumn{1}{c}{$E_{R}$}&\multicolumn{1}{c}{$\Gamma$}&
				\multicolumn{1}{c}{$E_{R}$}&\multicolumn{1}{c}{$\Gamma$}&\multicolumn{1}{c}{Partial wave}&
				\multicolumn{1}{c}{$E_{R}$}&\multicolumn{1}{c}{$\Gamma$}&	\multicolumn{1}{c}{$E_{R}$}&\multicolumn{1}{c}{$\Gamma$}\\
				\hline
				$S$&-0.0791821&2.40[4]&-0.07922&2.3[4] &$S$&-0.0765986&2.32[4]&-0.07654&2.3[4]~\cite{Ren2011}\\
				&-0.0688271&3.40[5] &-0.06888 &3.0[5] &   &&&-0.076285&2.08[4]~\cite{Strasburger2018}\\
				$P$&-0.0776826&1.17[4]&        &       &   &-0.0663292&1.46[4]&-0.06631&1.5[4]~\cite{Ren2011}\\
				$D$&-0.0748760&9.73[5]&        &       &	&&&-0.066186&1.47[4]~\cite{Strasburger2018}\\
				$F$&-0.0710955&1.25[4]&&&&-0.0635583&5.19[5]&-0.06356&5.0[5]~\cite{Ren2011}\\
				\multicolumn{5}{c}{He($1s3s$) threshold ($E_{t}=-0.0687695)$}&&&&-0.063446&6.67[5]~\cite{Strasburger2018}\\
				\cline{1-5}
				$S$&-0.0667958&1.15[3]&&&	&-0.0627822&1.96[5]&-0.06278&2.0[5]~\cite{Ren2011}\\
				&-0.0636442&2.30[4]   &&&&&&-0.062612&6.2[5]~\cite{Strasburger2018}\\
				&-0.0627965&5.72[5]   &-0.06279&8.0[5] & $P$&-0.0753283&1.69[4] &&   \\		
				$P$&-0.0643551&1.55[4]&&&	&-0.0656814&9.49[5]&&  \\
				&-0.0633909&1.71[4]&&&	&-0.0633338&4.24[5]&&   \\
				&-0.0627100&4.28[5]&&&&-0.0628784&8.11[5] &&   \\
				$D$&-0.0649959&2.71[4]&&&$D$&-0.0729422&1.46[4]&&    \\
				&-0.0630197&4.78[5]&&& &-0.0646616&5.36[5] &&  \\
				&-0.0625734&1.52[5]&&& &-0.0629700&1.62[5]&&\\	
				$F$&-0.0636423&3.80[5]&	&&$F$&-0.0695812&1.13[4]&&\\
				&-0.0626624&6.84[6]&&&&-0.0634685&2.20[5]&&\\		
				\textcolor{blue}{\textbf{$G$}}&	\textcolor{blue}{\textbf{-0.0667802}}&\textcolor{blue}{\textbf{1.39[4]}}&& &$G$&-0.0656173&6.80[5]&&\\
				&-0.0626398&7.56[6]&&&\multicolumn{5}{c}{Ps($n=2$) threshold ($E_{t}=-0.0625)$}\\
				\cline{6-10}
				\multicolumn{5}{c}{Ps($n=2$) threshold ($E_{t}=-0.0625)$}&\textcolor{blue}{\textbf{$D$}}&	\textcolor{blue}{\textbf{-0.0614237}}&	\textcolor{blue}{\textbf{4.13[4]}}&&\\	
				\cline{1-5}
				\textcolor{blue}{\textbf{$D$}}&\textcolor{blue}{\textbf{-0.0619721}}&\textcolor{blue}{\textbf{2.99[4]}}&&&\multicolumn{5}{c}{He($1s3s$) threshold ($E_{t}=-0.0612539)$}\\
				\cline{6-10}
				\multicolumn{5}{c}{He($1s3p$) threshold ($E_{t}=-0.0579844)$}&$S$&-0.0561153&4.64[6]&&\\
				\cline{1-5}
				$D$&-0.0579707&4.42[5]&&&$P$&-0.0563195&4.84[5]\\
				\multicolumn{5}{c}{He($1s3d$) threshold ($E_{t}=-0.0556585)$}&$D$&-0.0561555&5.02[5]\\
				\cline{1-5}
				$S$&-0.0366885&1.73[5]&&&\multicolumn{5}{c}{He($1s3p$) threshold ($E_{t}=-0.0560323)$}\\	
				\cline{6-10}
				$P$&-0.0398059&4.50[5]&&&$S$&-0.0390939&4.66[6]&-0.038256&3.8[5]~\cite{Strasburger2018}\\
				&-0.0365834&1.12[5]&&&&-0.0346531&4.48[5]&&\\
				$D$&-0.0391985&4.07[5]&&&$P$&-0.0381070&1.10[4]&&\\
				\multicolumn{5}{c}{He($1s4s$) threshold ($E_{t}=-0.0365567)$}&	&-0.0340958&3.78[5]\\
				\cline{1-5}	
				$S$	&-0.0325098&2.43[5]& &&  &-0.0340231&4.85[5]&&\\	
				$P$&-0.0327085&1.93[5]&& &$D$&-0.0382568&5.30[5]&&\\
				&-0.0323252&2.40[5] &&&   &-0.0350187&5.98[5]&&\\
				$D$&-0.0353585&1.05[4] &&&   &-0.0341024&2.04[5]&&\\	
				&-0.0324918&2.67[5] &&&$F$&-0.0374726&6.13[5]&&\\
				\multicolumn{5}{c}{He($1s4p$) threshold ($E_{t}=-0.0322837)$}&&-0.0341508&6.73[5]	\\
				\cline{1-5}	
				&& &&&   &-0.0336354&1.40[5]&&\\
				&&	&&&$G$&-0.0363999&8.05[5]&&\\
				&&&&&\multicolumn{5}{c}{He($1s4s$) threshold ($E_{t}=-0.0335957)$}\\	
			\end{tabular}
			%}
		\end{ruledtabular}
	\end{table*}
}
\subsubsection{Resonant structures in Ps-formation cross sections}

In addition to the narrow Feshbach resonances discussed above, we find
several resonant structures with widths of order $10^{-4}$ a.u.. These resonances are indicated in bold in Table~\ref{t2}. A distinctive feature
of these states is that they produce pronounced structures in the
Ps-formation cross sections, although the corresponding eigenphase sums
do not show the usual full resonant increase of approximately $\pi$
radians.

\begin{figure*}[htbp]
	\centering
	\subfigure{
		\includegraphics[scale=0.3]{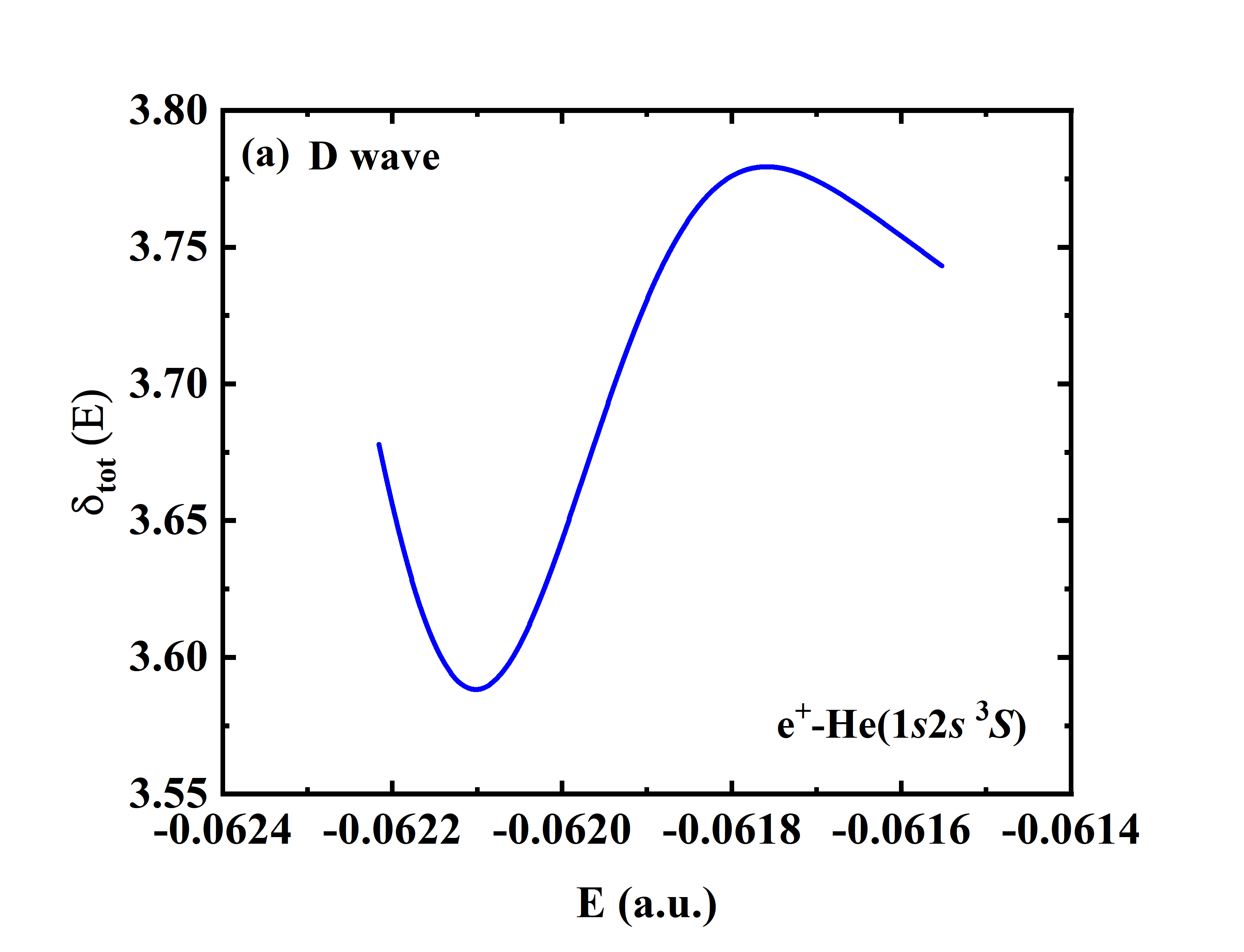}
		\label{fig5a}
	}
	\subfigure{
		\includegraphics[scale=0.3]{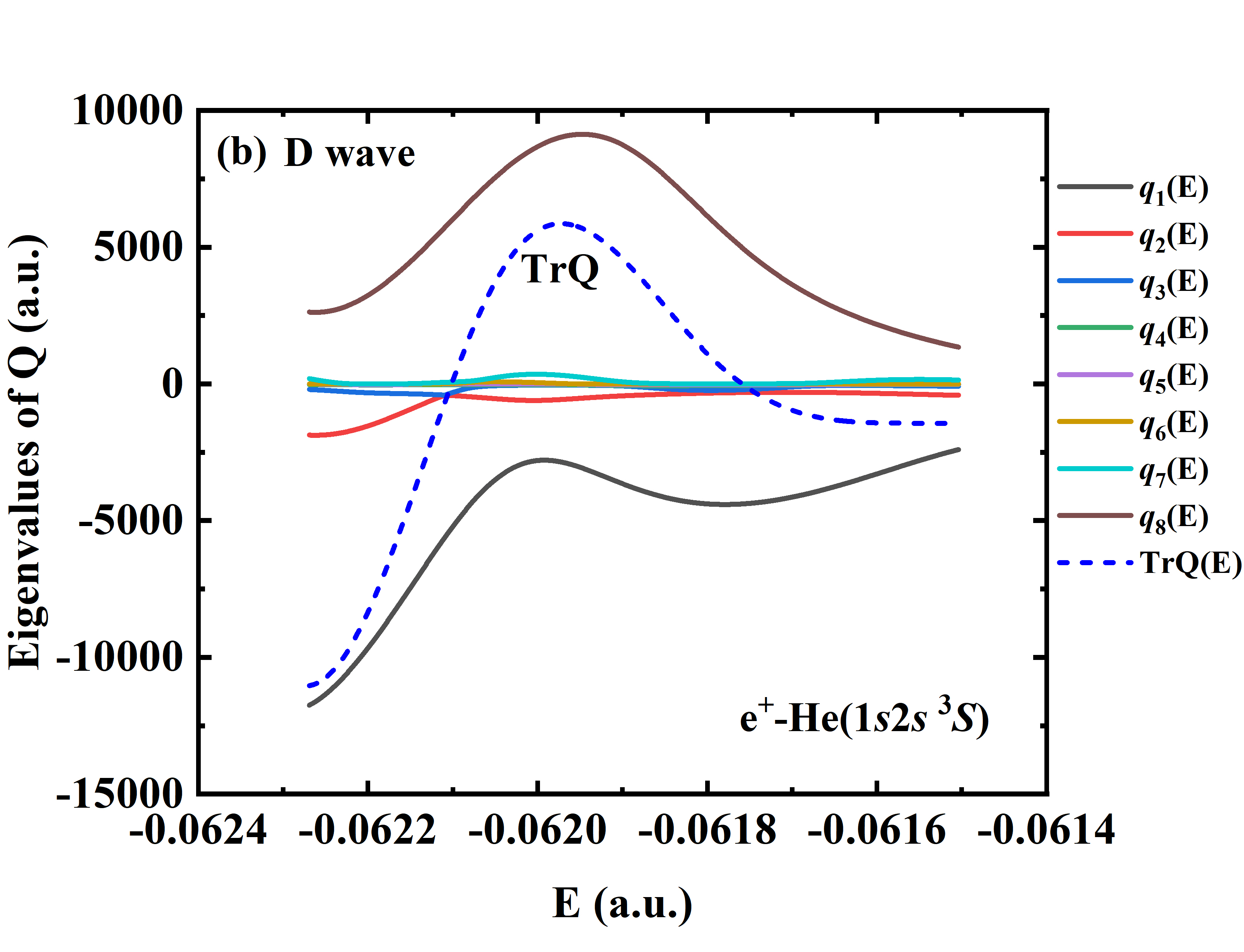}
		\label{fig5b}
	}
	\caption{(Color online) (a) $D$-wave eigenphase sums for positron scattering from metastable He($1s2s\,^3 S$). (b) Eigenvalues of the $D$-wave time-delay matrix $Q(E)$ (full curves) and ${\rm Tr}\,Q(E)=2\hbar (d\delta_{\rm tot}(E)/{dE})$ (broken curve).}
	\label{fig5}
\end{figure*}

Figure~\ref{fig5a} shows the $D$-wave eigenphase sum for the
$e^{\scriptscriptstyle +}$--$\mathrm{He}(1s2s\,^{3}S)$ system in the
energy region between the Ps($n=2$) and $\mathrm{He}(1s3p)$ thresholds.
Only a modest increase of the eigenphase sum is observed in this region.
To further characterize this structure, Fig.~\ref{fig5b} shows the
time-delay eigenvalues $q_i(E)$, together with ${\rm Tr}\,Q$, for
$-0.0624 \leq E \leq -0.0614$ a.u. There are eight open channels in this
energy interval. Among the time-delay eigenvalues, $q_8$ displays a
clear peak near $E=-0.0619$ a.u., whereas the remaining eigenvalues vary
only slowly with energy. A fit over an energy interval
$\Delta E=0.001$ a.u. around the peak gives
$E_r=-0.0619721$ a.u. and $\Gamma=3.0\times10^{-4}$ a.u. for the
$D$-wave resonance, with a large negative background parameter
$C=-3320$.

This behavior is similar to that reported in Ref.~\cite{Igarashi2004}, where a
rapidly decreasing background eigenphase gives rise to a large negative
background contribution in the time-delay signal. As a result, the net
increase of the eigenphase sum is strongly reduced, and no full
$\pi$-radian rise is observed, even though a time-delay eigenvalue
exhibits a clear resonance-like peak. These states are therefore more
clearly identified from the time-delay analysis than from the eigenphase
sum alone.

\begin{figure*}[htbp]
	\centering
	\subfigure{
		\includegraphics[scale=0.3]{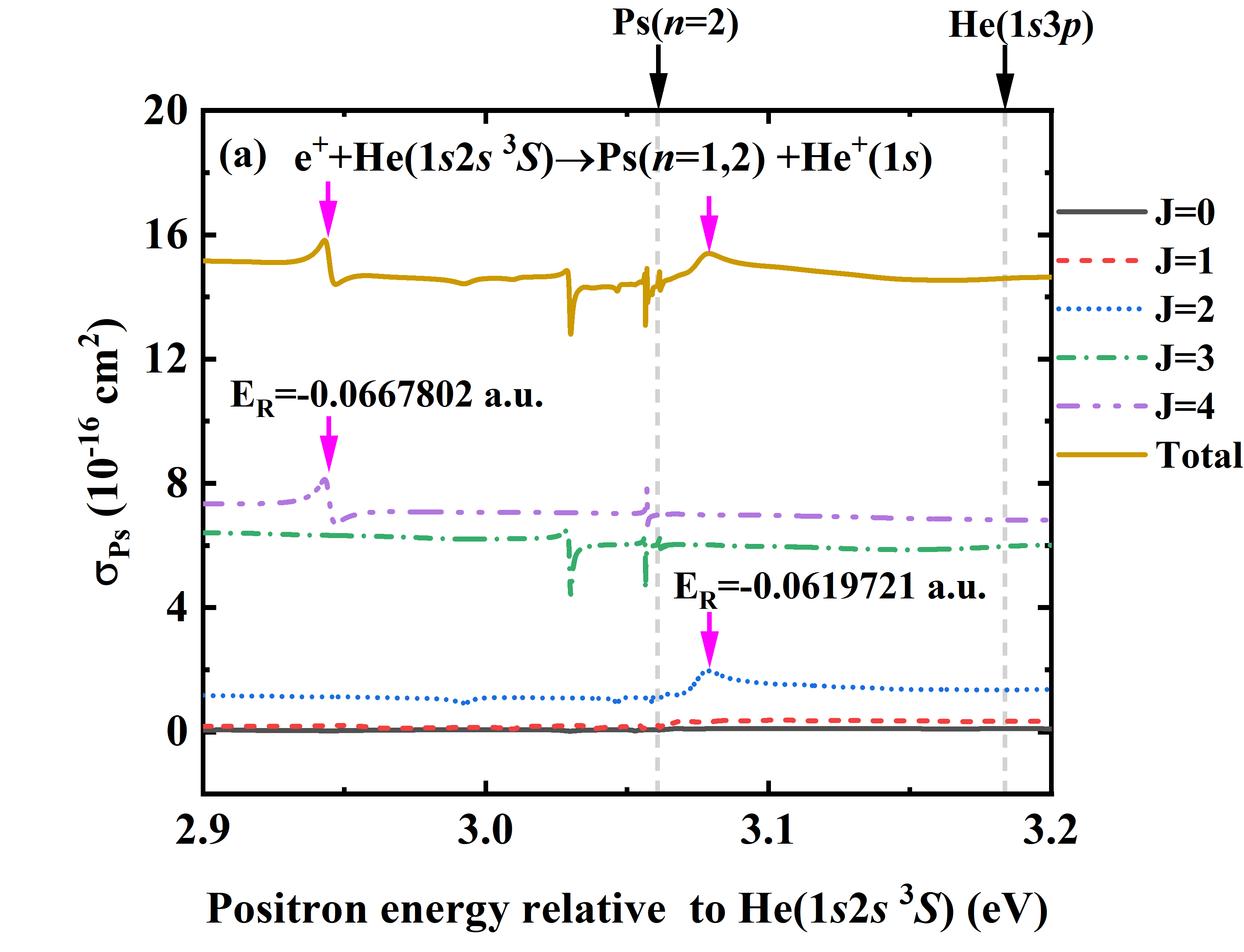}
		\label{fig6a}
	}
	%\quad
	\subfigure{
		\includegraphics[scale=0.3]{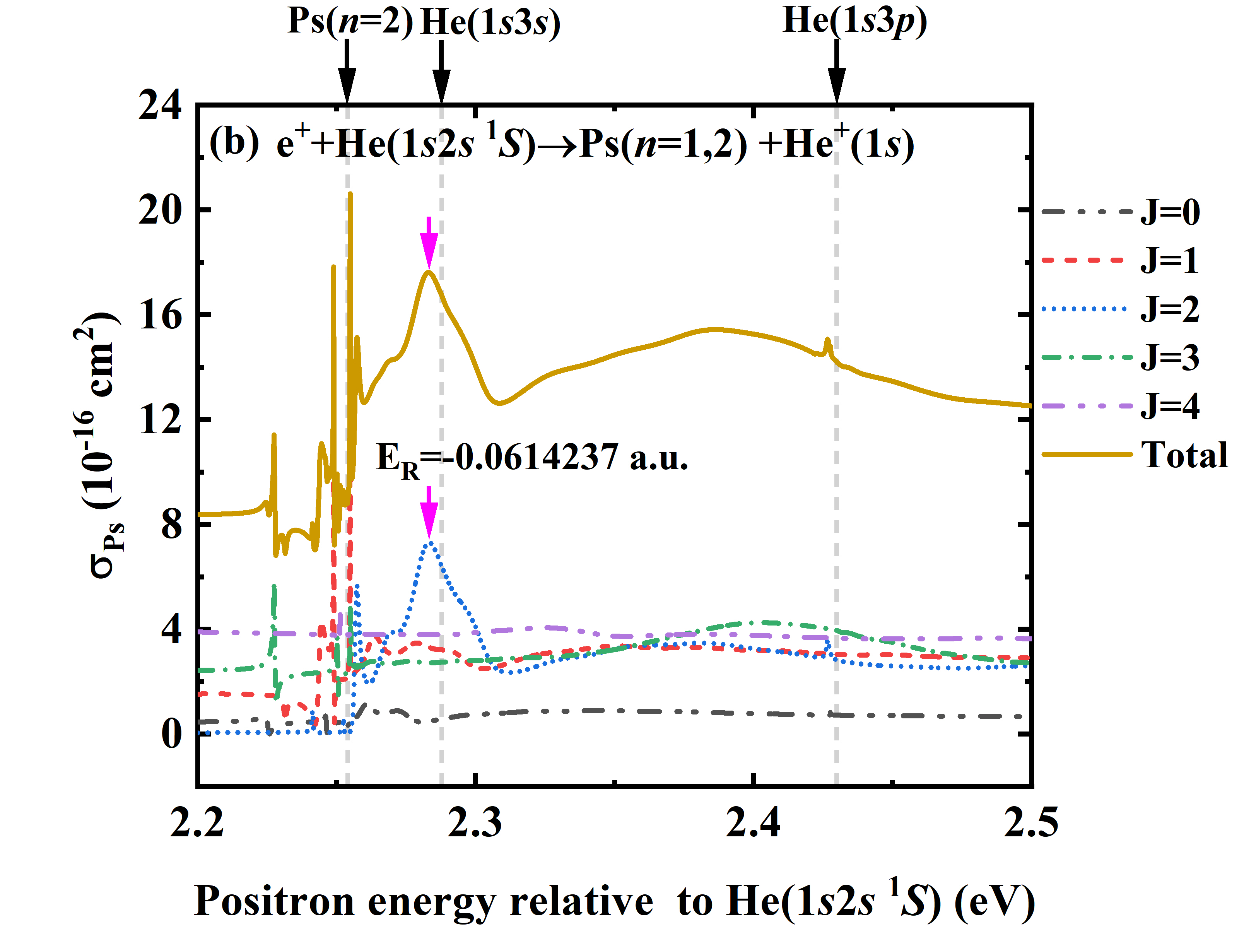}
		\label{fig6b}
	}
	\caption{(Color online) (a) Characteristic structures in the partial and total Ps formation cross sections for the $e^{\scriptscriptstyle+}$-He($1s2s\,^{3}S$) system; (b) same as (a) for the $e^{\scriptscriptstyle+}$-He($1s2s\,^{1}S$) system. Pink arrows indicate resonance positions, and gray vertical dashed lines denote the Ps($n=2$), He($1s3s$), and He($1s3p$) thresholds.}
	\label{fig6}
\end{figure*}

For the $e^{\scriptscriptstyle +}$--$\mathrm{He}(1s2s\,^{3}S)$ system,
we find two such resonances that give rise to visible structures in the
Ps-formation cross section. These structures are shown in
Fig.~\ref{fig6a}, together with the contributions from individual partial
waves. The lower-energy structure, near $E_r=-0.0667802$ a.u., is mainly
associated with a $G$-wave resonance. The second structure, near
$E_r=-0.0619721$ a.u., is dominated by the $D$-wave resonance discussed
above.

For the $e^{\scriptscriptstyle +}$--$\mathrm{He}(1s2s\,^{1}S)$ system,
we find one such obvius resonant structure in the Ps-formation cross
section, as shown in Fig.~\ref{fig6b}. The structure occurs near
$E_r=-0.0614237$ a.u. and is mainly associated with a $D$-wave
resonance. In the partial-wave-resolved cross sections, the $J=2$
contribution exhibits a distinct enhancement in the same energy region.
The structure lies close to the $\mathrm{He}(1s3s)$ threshold and above
the Ps($n=2$) threshold, where the coupling between the helium-excitation channels and the Ps-formation channels is strong.

\section{Summary}
\label{sec:summary}

Near-threshold resonances have been analyzed using both the eigenphase
sum and the time-delay matrix. We find narrow Feshbach resonance series
in higher partial waves extending up to highly excited atomic thresholds in both the singlet and triplet systems.
In addition, several resonances with widths of order $10^{-4}$ a.u. are
identified from clear peaks in the time-delay eigenvalues. For these
states, the eigenphase sums do not exhibit a full $\pi$-radian increase
because of the strongly varying background. Nevertheless, they generate
visible structures in the Ps-formation cross sections.

In the singlet
system, a mainly $D$-wave resonance near the $\mathrm{He}(1s3s)$
threshold produces a structure around 2.3 eV. In the triplet system,
structures in the Ps($n=2$) threshold region are associated mainly with
$G$- and $D$-wave resonances. These resonance-induced structures, with
observed widths of several tens of meV in the total Ps-formation cross
sections, may provide useful signatures for positron attachment to metastable
helium.
Finally, it should be noted that the present three-body model relies on the frozen-core approximation. Since He$^{\scriptscriptstyle+}$ is an open-shell ion, the validity of this approximation for this system and its possible influence on the calculated cross sections should be tested in future work.

\begin{acknowledgments}
	
This work is supported by the National Natural Science Foundation of China under Grant Nos. 12374235, 12274423, 12393821 and 12174402; and by the Pioneer Research Project for Basic and Interdisciplinary Frontiers of Chinese Academy of Sciences under Grants No. XDB0920101 and No. XDB0920100. All the calculations are done on the APM-Theoretical Computing Cluster(APM-TCC).

\end{acknowledgments}

\section*{Data Availability}
	
The data supporting the findings of this article have been tabulated with in the article. Additional metadata are available from the corresponding author upon request.

	\section*{Appendix: Dipole resonances series below the Ps($n=2$) threshold.}
%\label{sec:Appendix}
This Appendix presents a series of dipole resonances below the Ps($n=2$) threshold. The energies of Ps atom are degenerate with respect to the $l$ quantum number, and hence, the Ps-A$^{\scriptscriptstyle+}$ interaction give a dipole interaction, i.e., an interaction-potential with long-range form proportional to $-1/r^2$. This long-range potential gives, in principle, an infinite sequence of quasibound states clustering towards the Ps($n=2$) thresholds. The resonance energies within each dipole series can be fitted by the following linear relation:
\begin{align}
	\label{lnEv}
	\ln(E_{\nu})=\ln(E_0)-\alpha\nu\,.
\end{align}

Table~\ref{t3} lists the energy ratios of successive resonances identified in the lowest four partial waves near the Ps($n=2$) threshold for the e$^{\scriptscriptstyle+}$--He($1s2s\,^{1,3}S$) systems. The fitted universal scaling parameters $\alpha$ are also listed.

\begin{figure*}[htbp]
	\centering
	\label{fig7}
	\renewcommand{\thesubfigure}{(a\arabic{subfigure})}
\setcounter{subfigure}{0}
\subfigure{
	\includegraphics[width=0.315\textwidth]{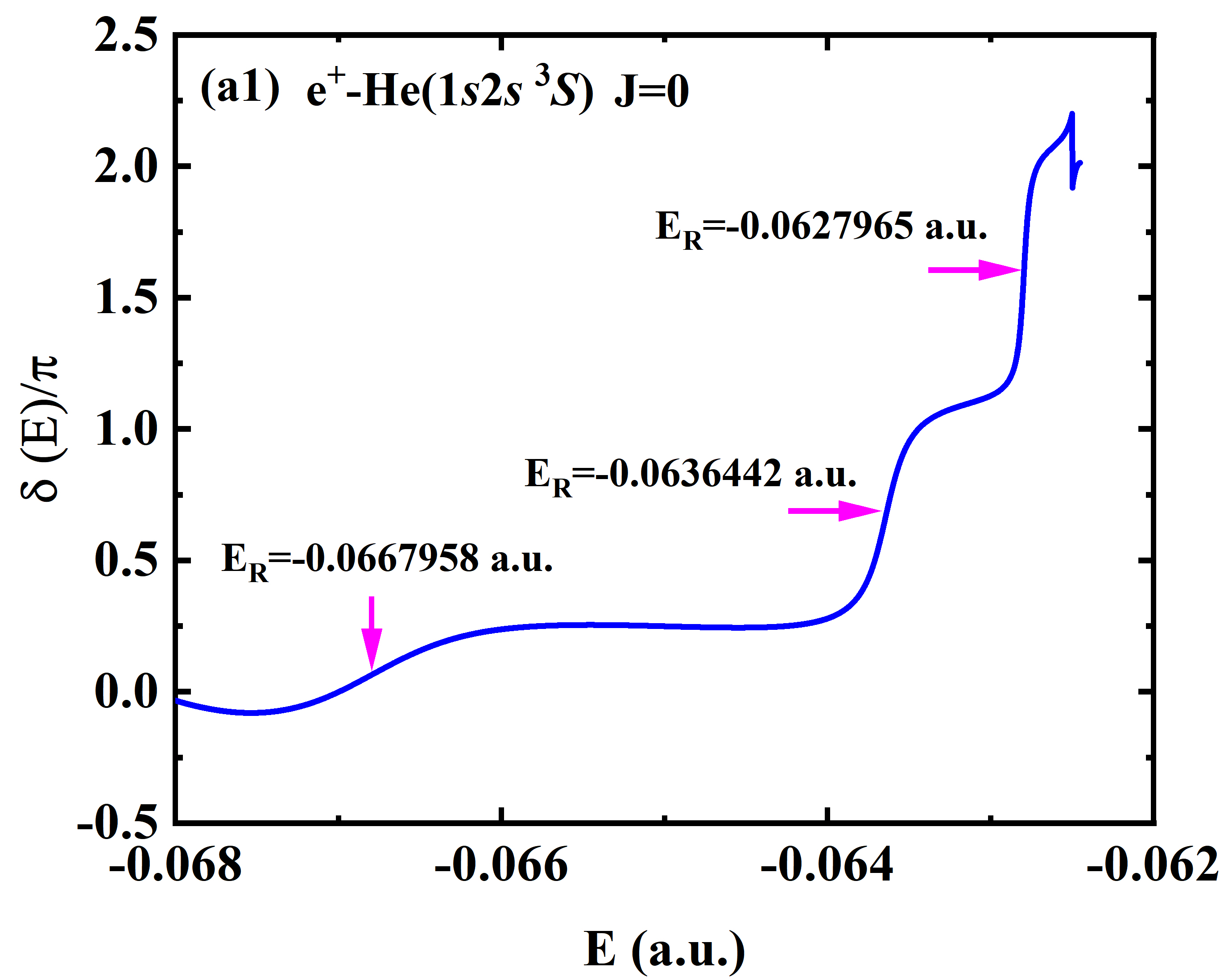}
	\label{fig7a1}
}
\subfigure{
	\includegraphics[width=0.305\textwidth]{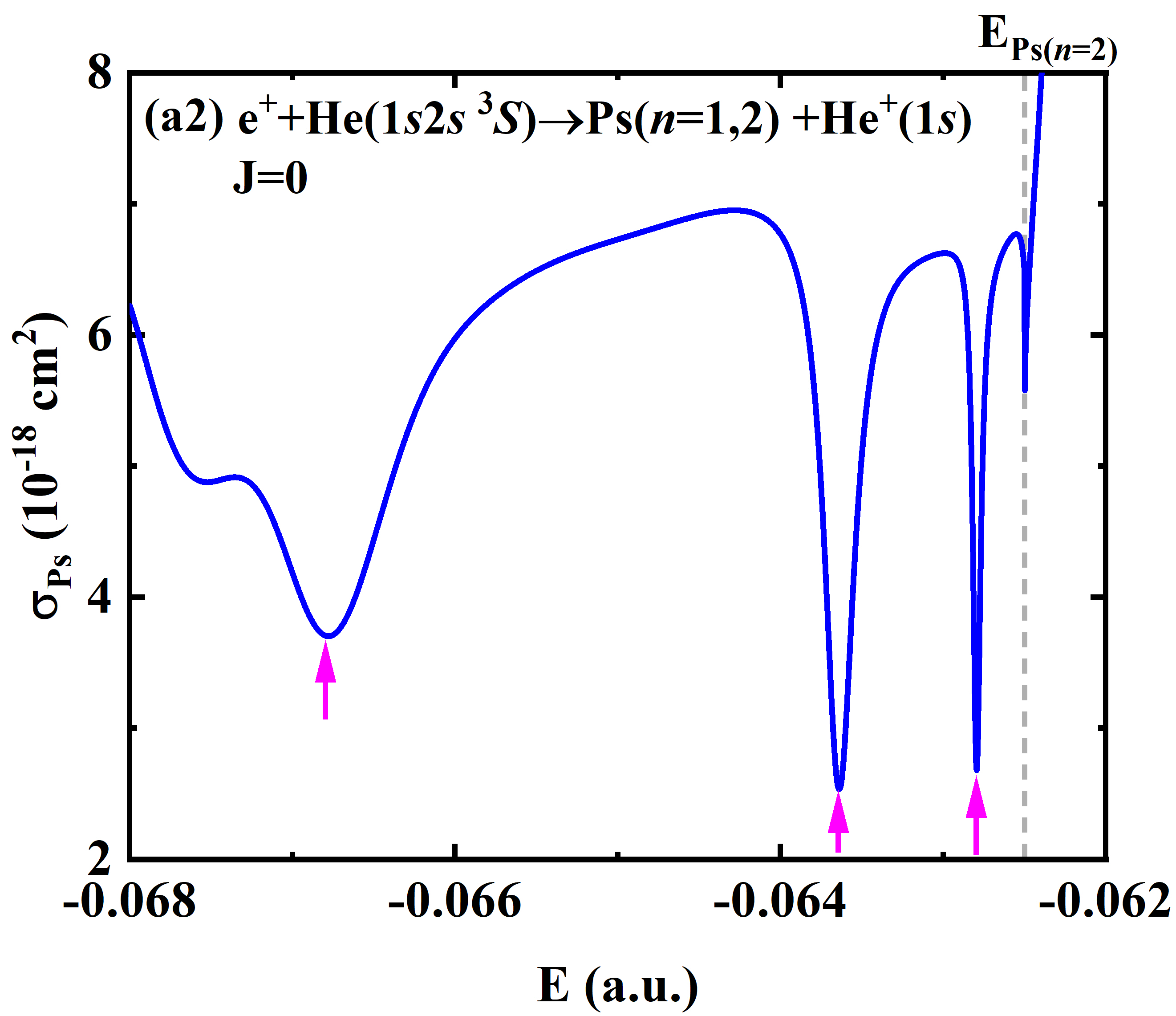}
	\label{fig7a2}
}
\subfigure{
	\includegraphics[width=0.30\textwidth]{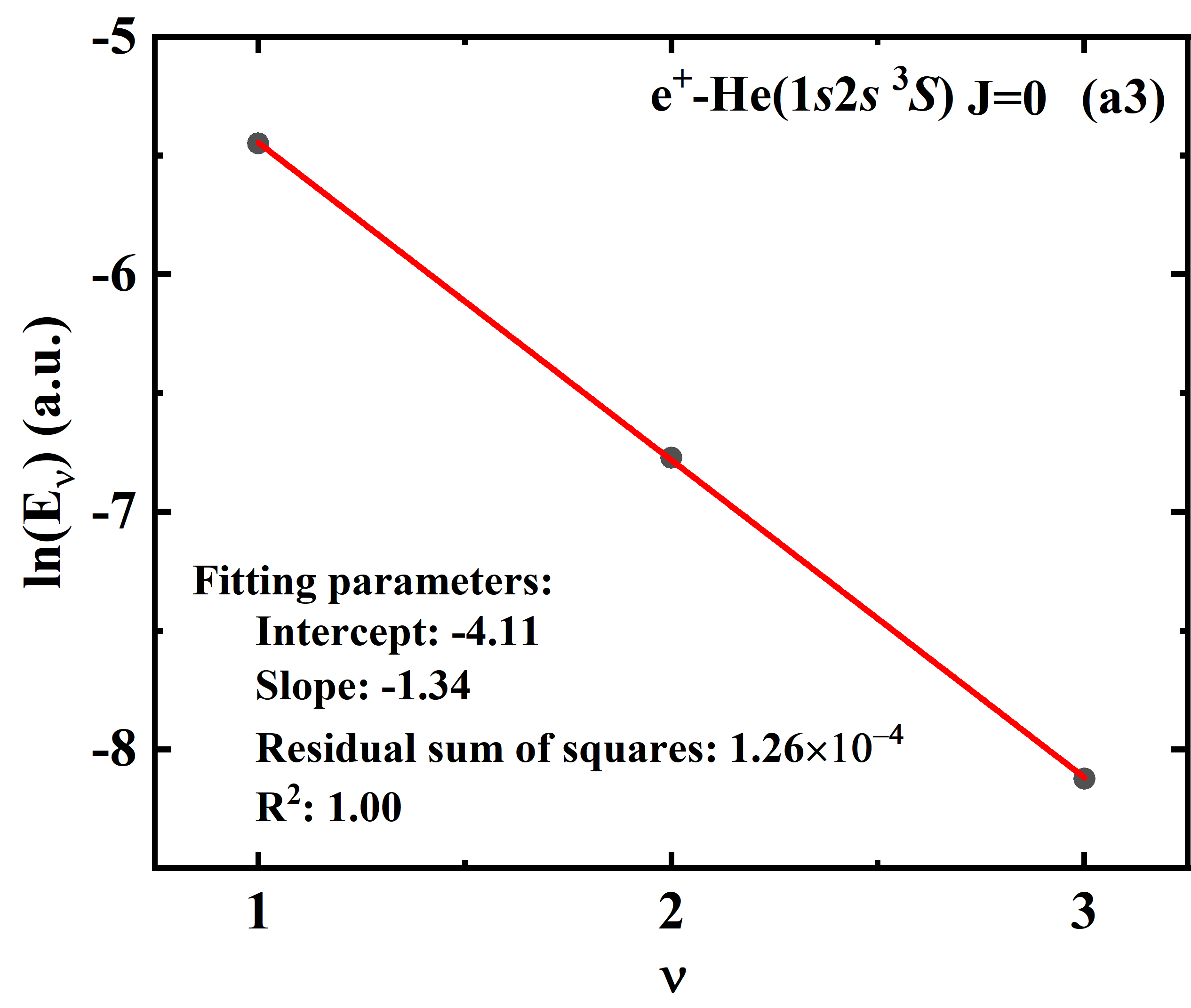}
	\label{fig7a3}
}\\
	\renewcommand{\thesubfigure}{(b\arabic{subfigure})}
\setcounter{subfigure}{0}
\subfigure{
	\includegraphics[width=0.31\textwidth]{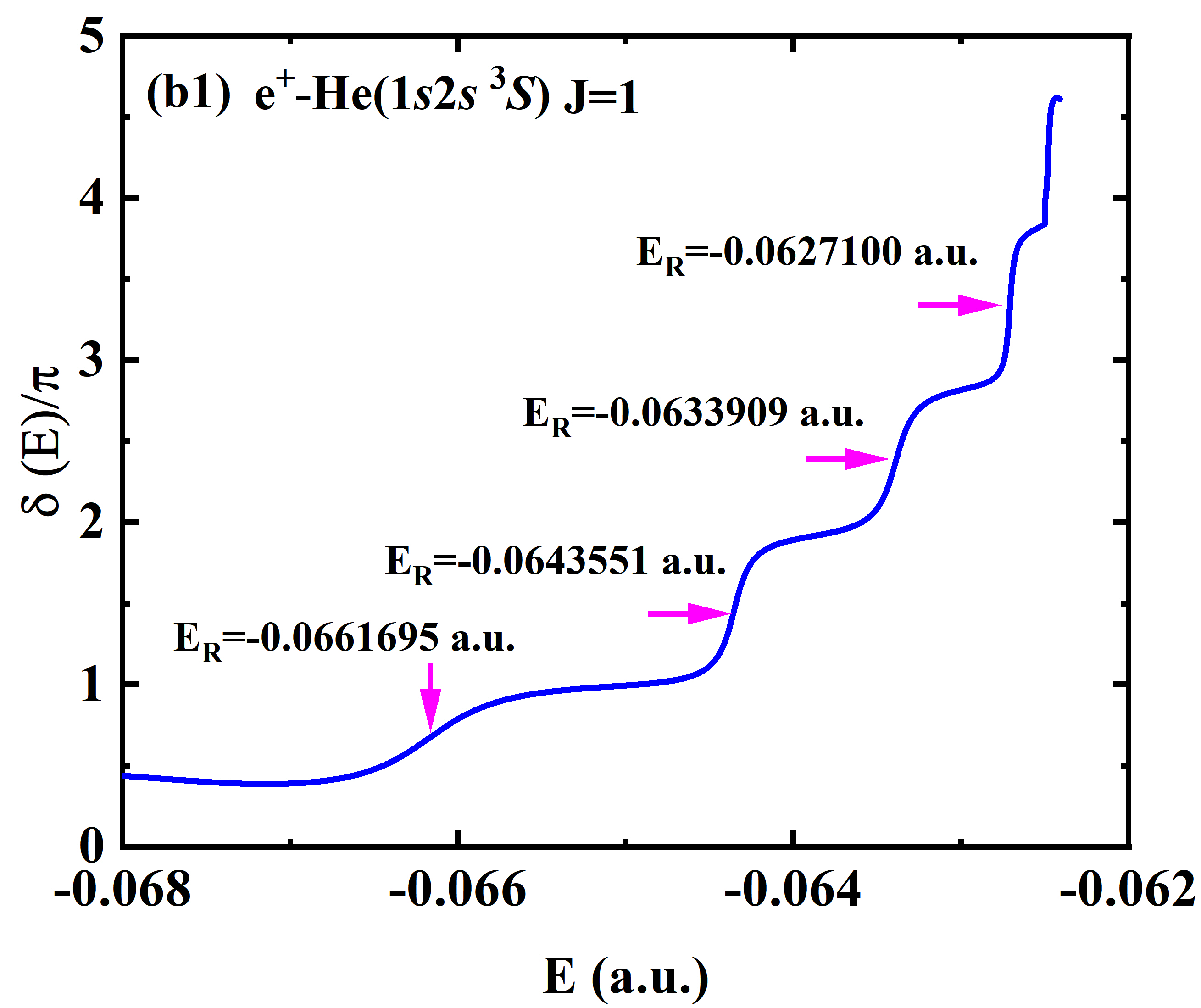}
	\label{fig7b1}
}
\subfigure{
	\includegraphics[width=0.32\textwidth]{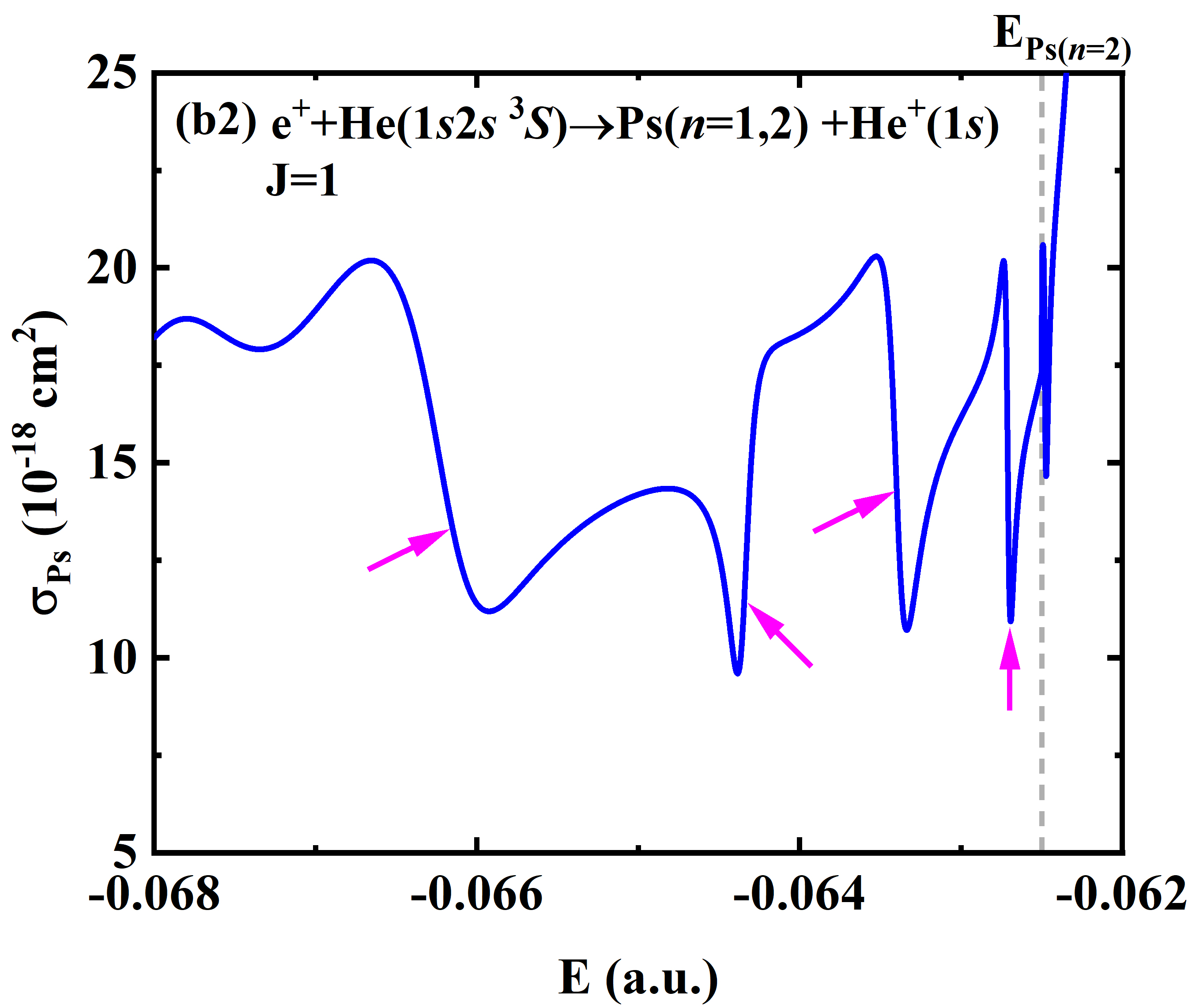}
	\label{fig7b2}
}
\subfigure{
	\includegraphics[width=0.31\textwidth]{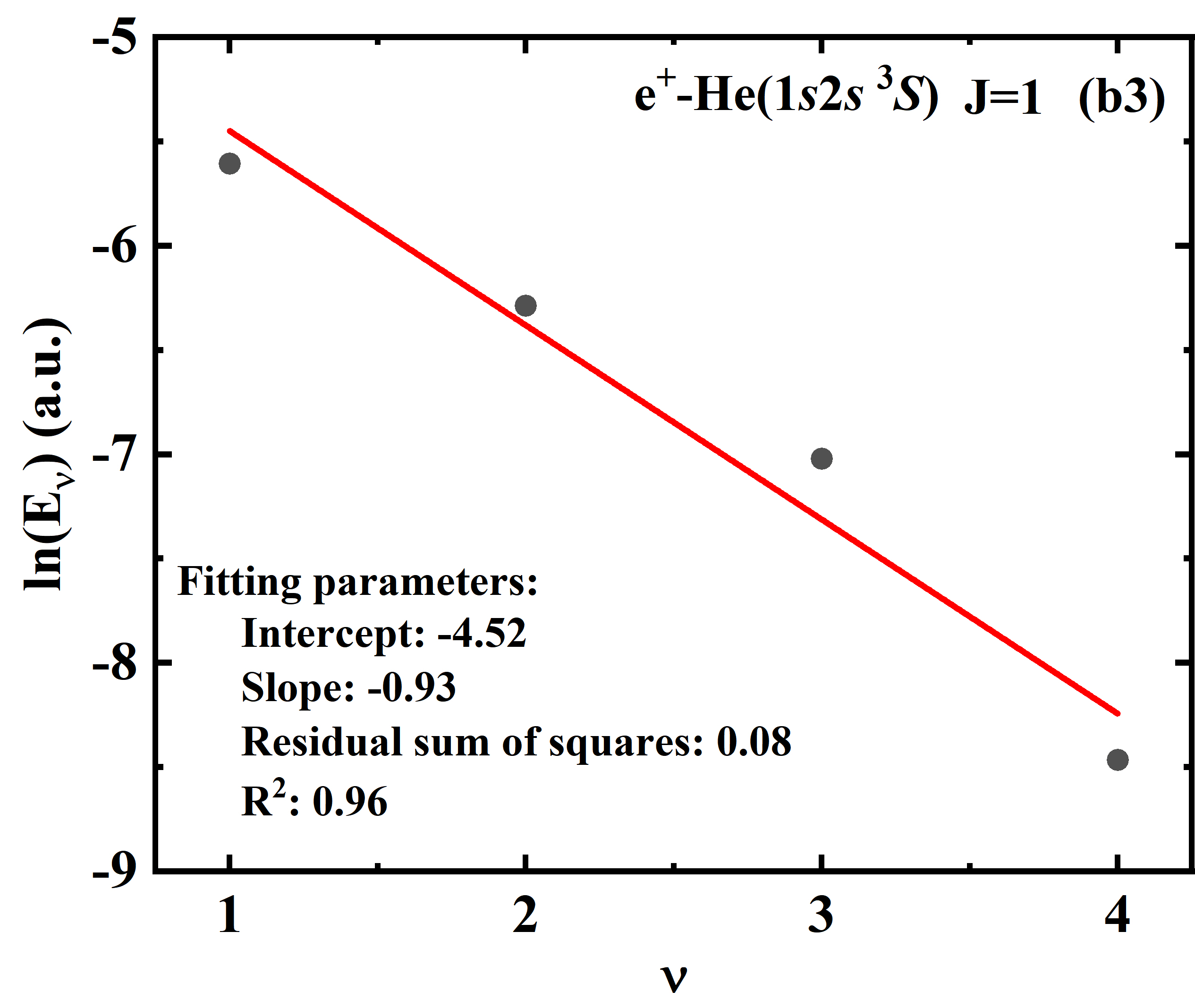}
	\label{fig7b3}
}\\
	\renewcommand{\thesubfigure}{(c\arabic{subfigure})}
\setcounter{subfigure}{0}
\subfigure{
	\includegraphics[width=0.318\textwidth]{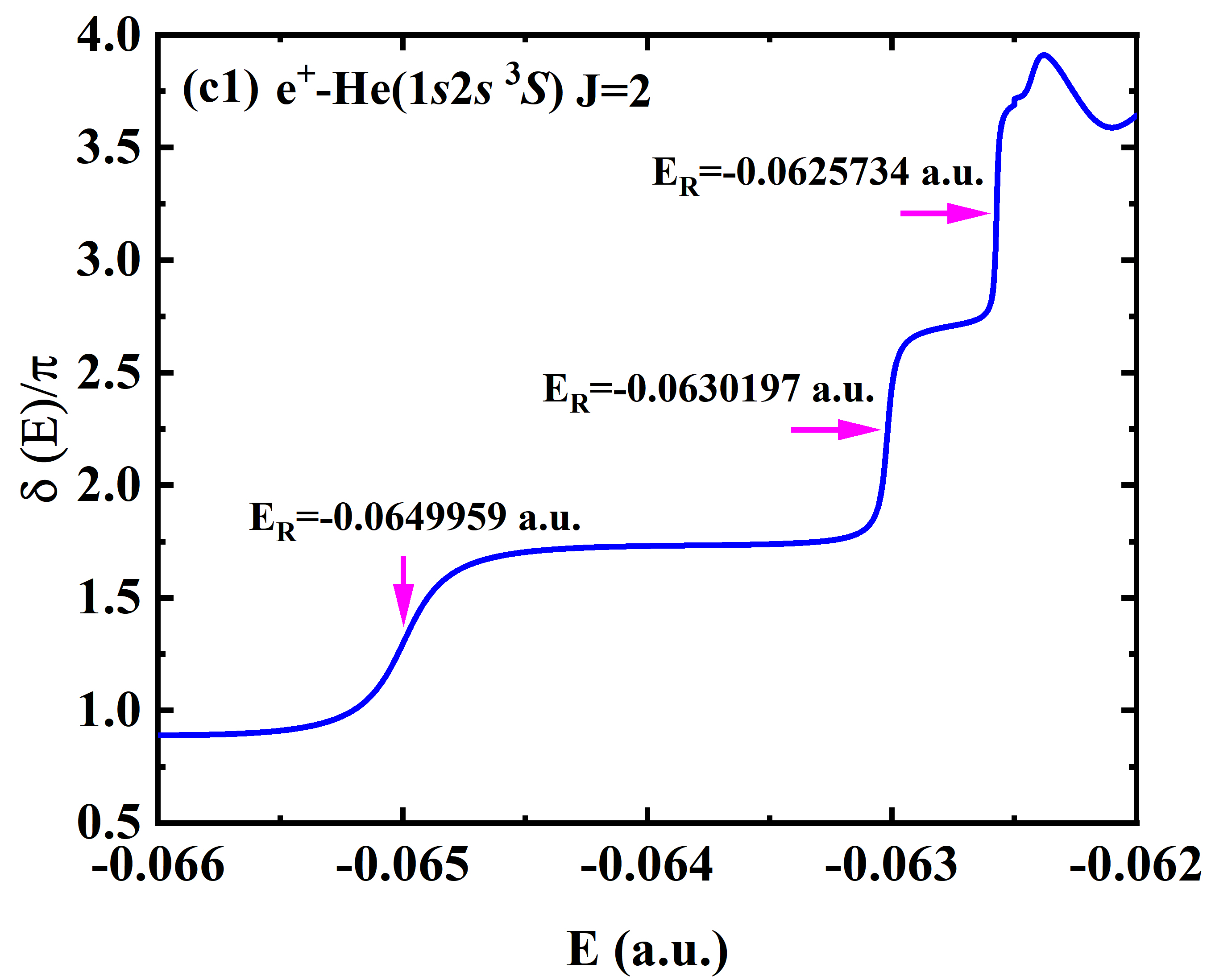}
	\label{fig7c1}
}
\subfigure{
	\includegraphics[width=0.318\textwidth]{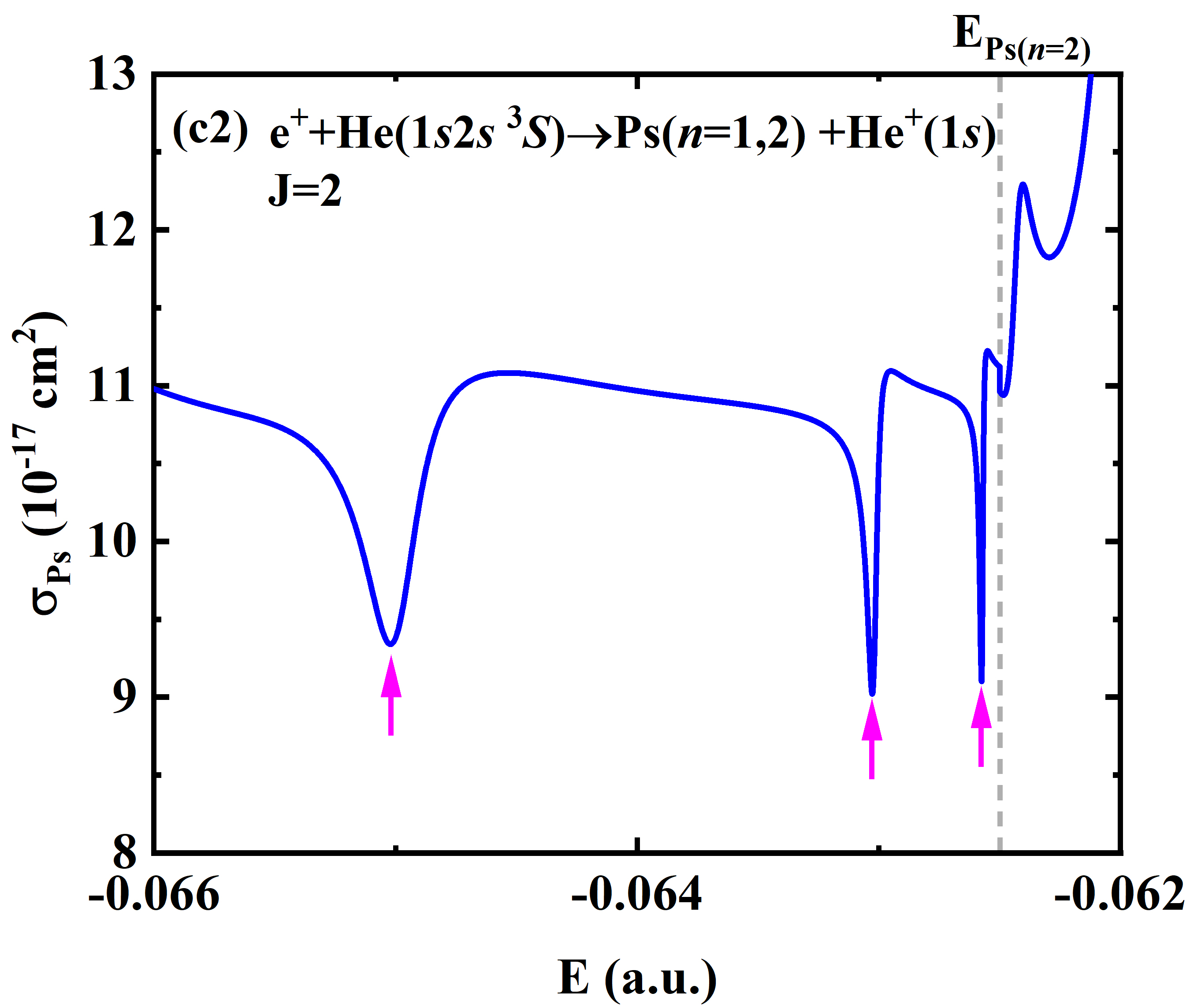}
	\label{fig7c2}
}
\subfigure{
	\includegraphics[width=0.315\textwidth]{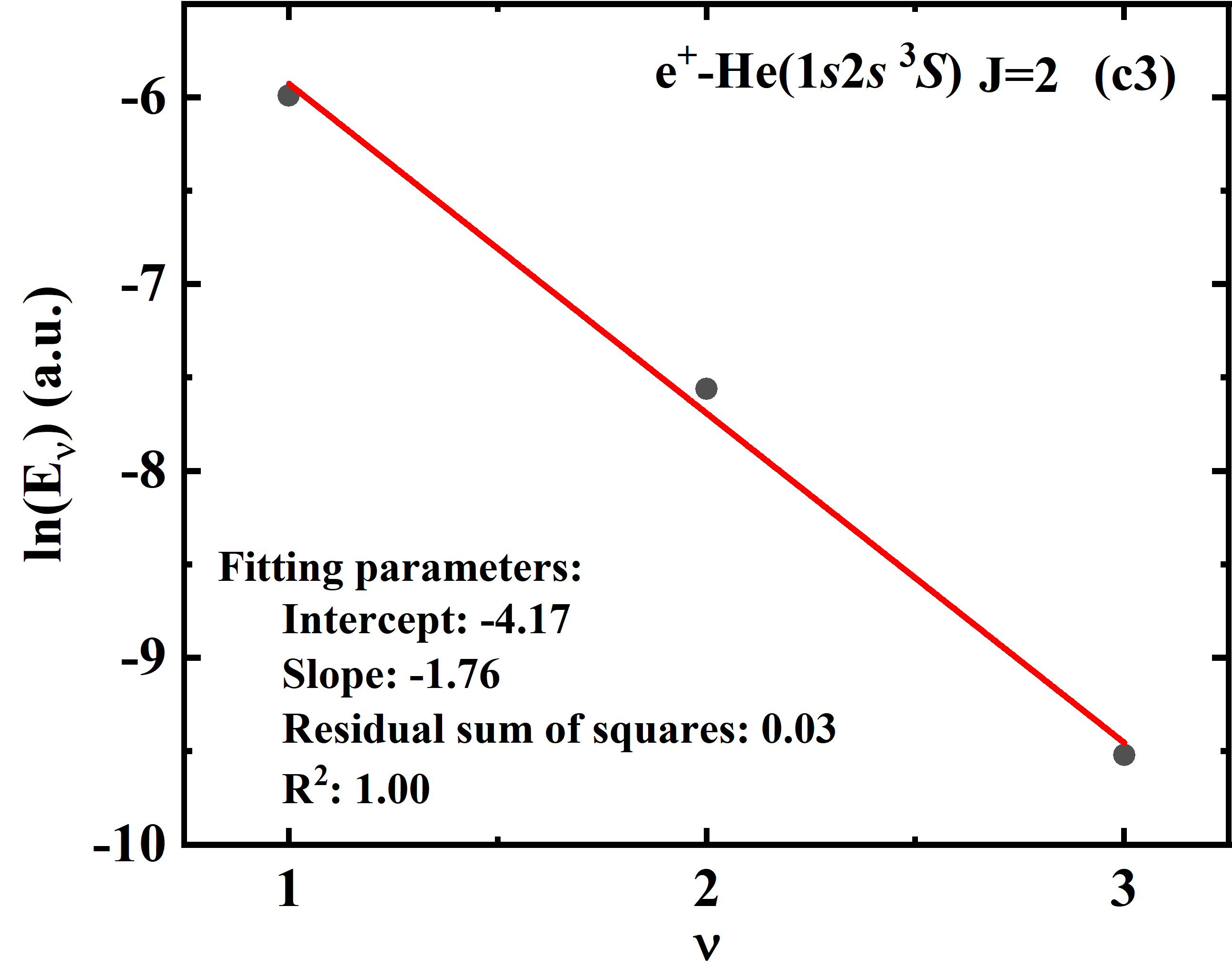}
	\label{fig7c3}
}
	\caption{(Color online) (a1), (b1), and (c1) Eigenphase sums as functions of energy; (a2), (b2), and (c2) Ps-formation cross sections for the e$^{\scriptscriptstyle+}$-He($1s2s\,^3S$) system with $J=0-2$ near the Ps($n=2$) threshold. The dashed lines indicate the Ps($n=2$) threshold and the arrows indicate the resonance positions. (a3), (b3), and (c3) Semilogarithmic plots of the resonance positions $E_{R}=E_\text{th}-E_{\nu}$ for Ps($n=2$)+He$^{\scriptscriptstyle+}(1s)$ below the Ps($n=2$) threshold. Straight lines represent the fits using Eq.~(\ref{lnEv}).}
\end{figure*}

\begin{figure*}[htbp]
	\centering
	\label{fig8}
	\renewcommand{\thesubfigure}{(a\arabic{subfigure})}
	\setcounter{subfigure}{0}
	\subfigure{
		\includegraphics[width=0.305\textwidth]{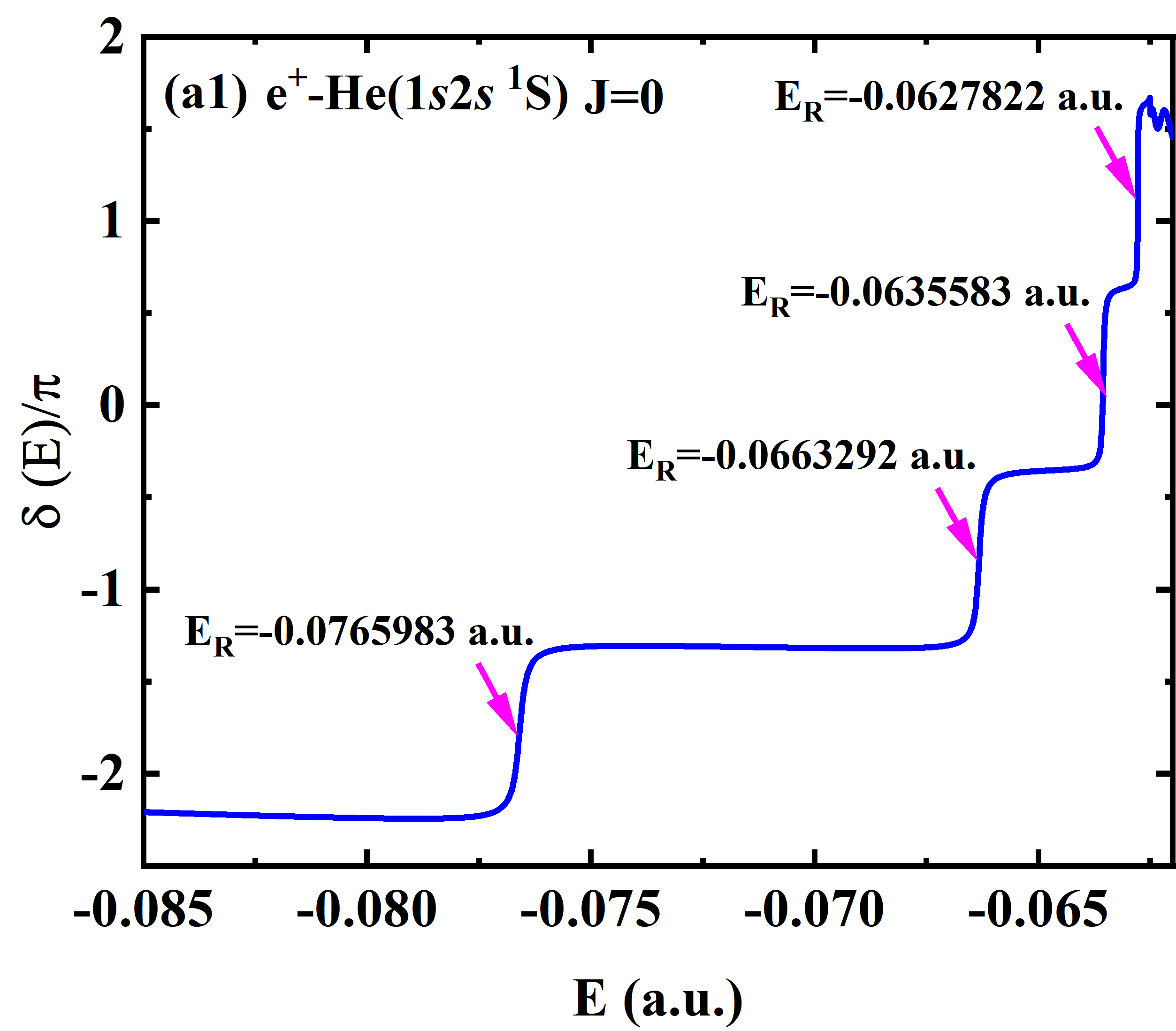}
		\label{fig8a1}
	}
	\subfigure{
		\includegraphics[width=0.33\textwidth]{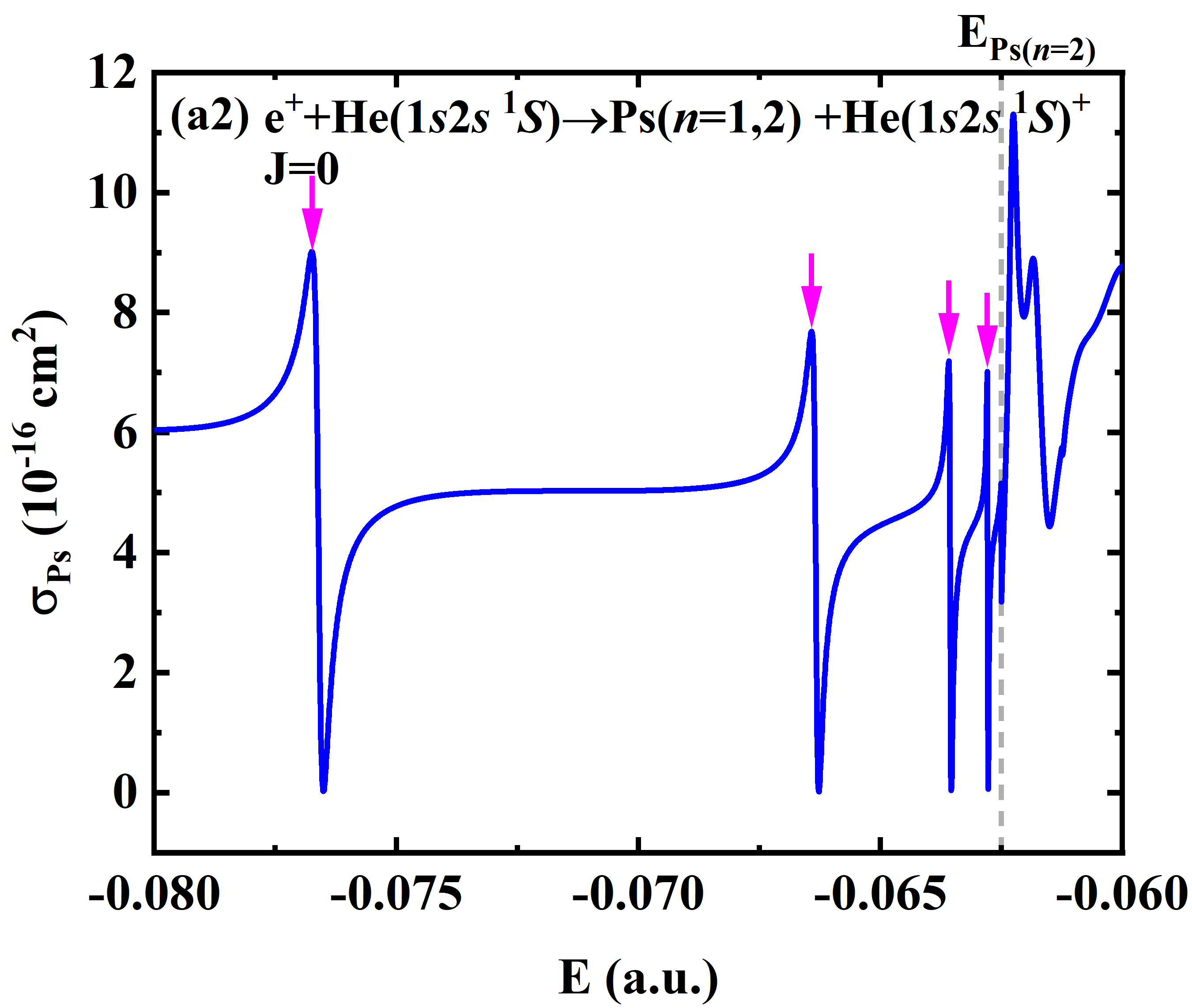}
		\label{fig8a2}
	}
	\subfigure{
		\includegraphics[width=0.315\textwidth]{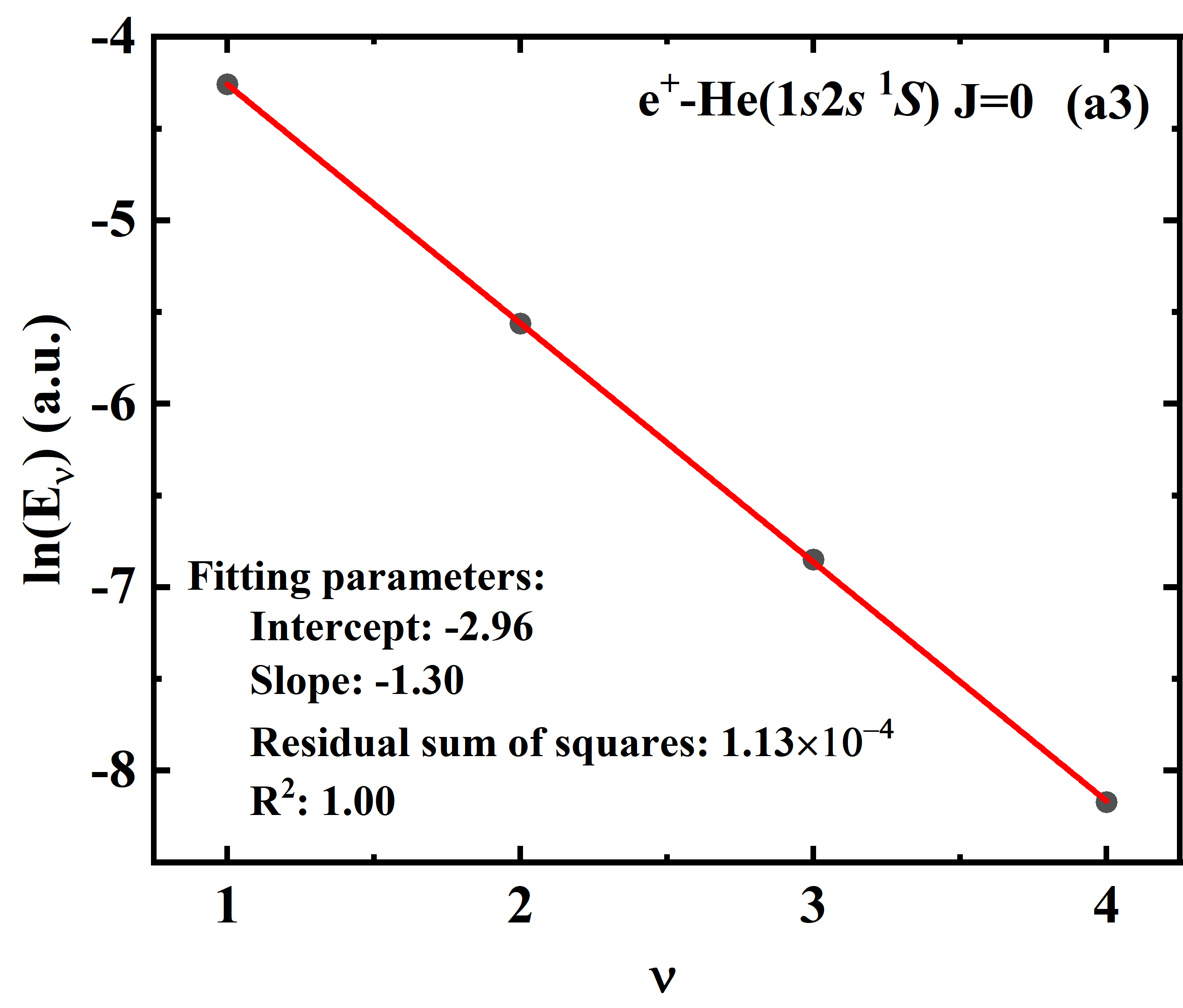}
		\label{fig8a3}
	}\\
	\renewcommand{\thesubfigure}{(b\arabic{subfigure})}
	\setcounter{subfigure}{0}
	\subfigure{
		\includegraphics[width=0.305\textwidth]{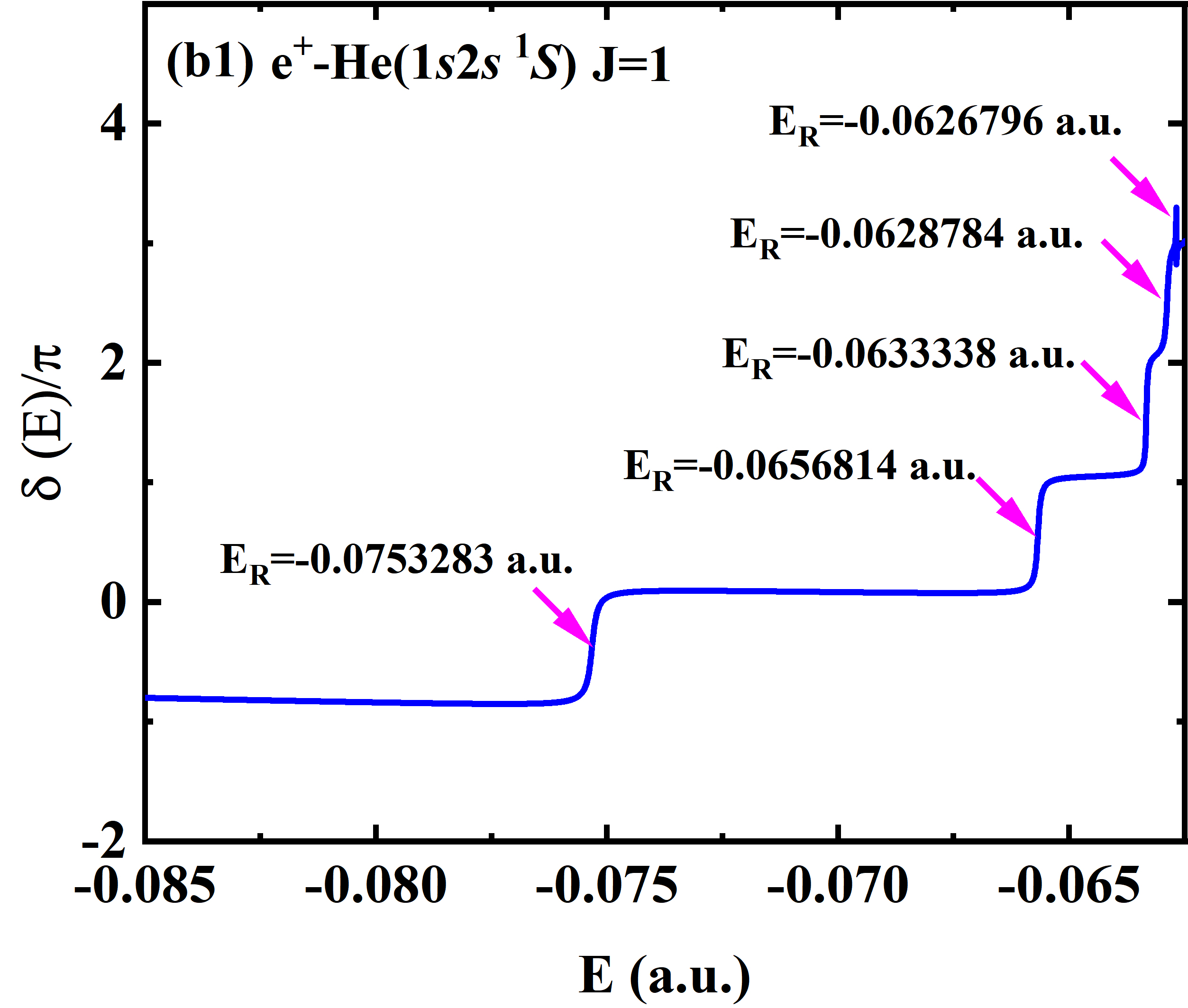}
		\label{fig8b1}
	}
	\subfigure{
		\includegraphics[width=0.33\textwidth]{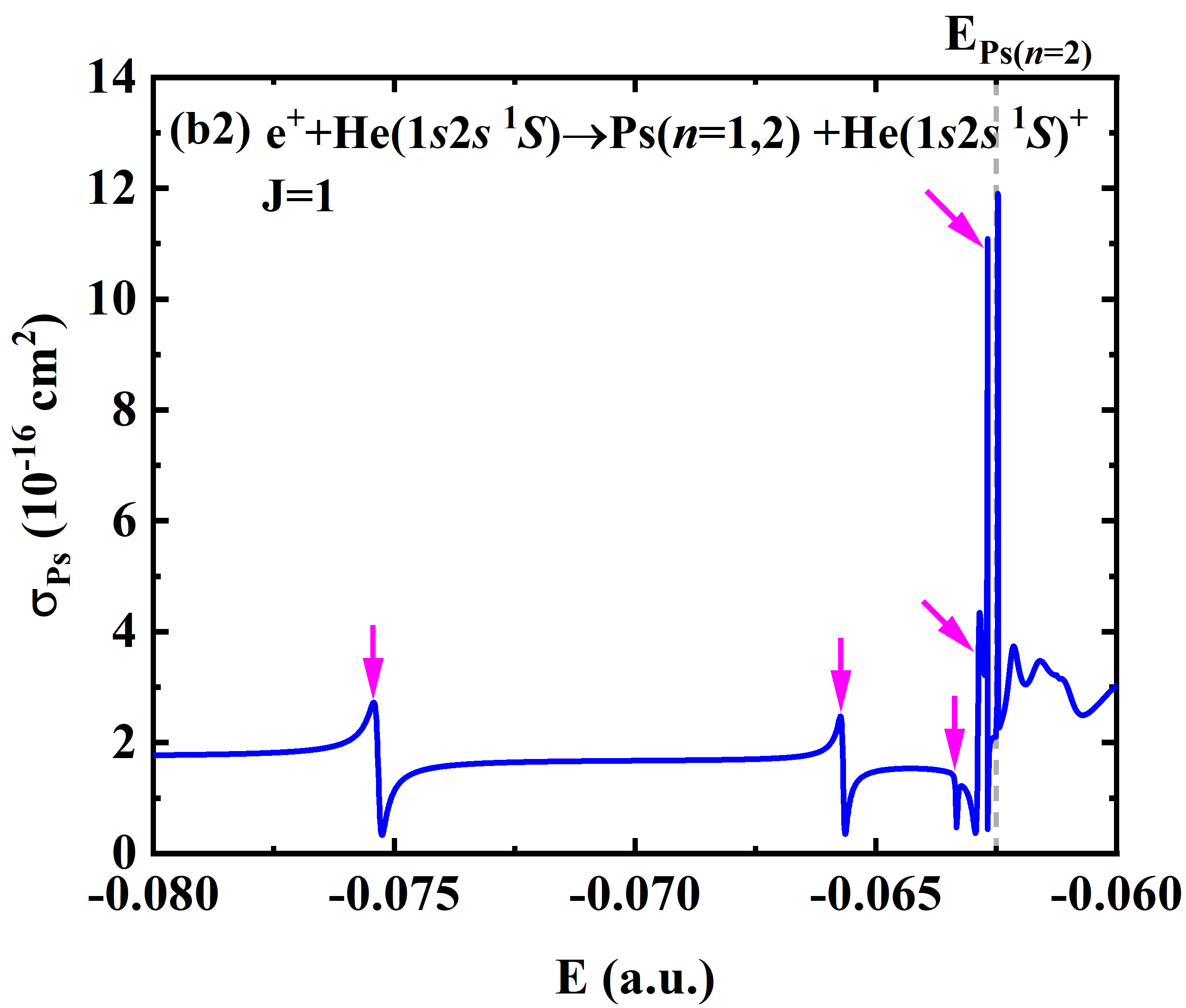}
		\label{fig8b2}
	}
	\subfigure{
		\includegraphics[width=0.315\textwidth]{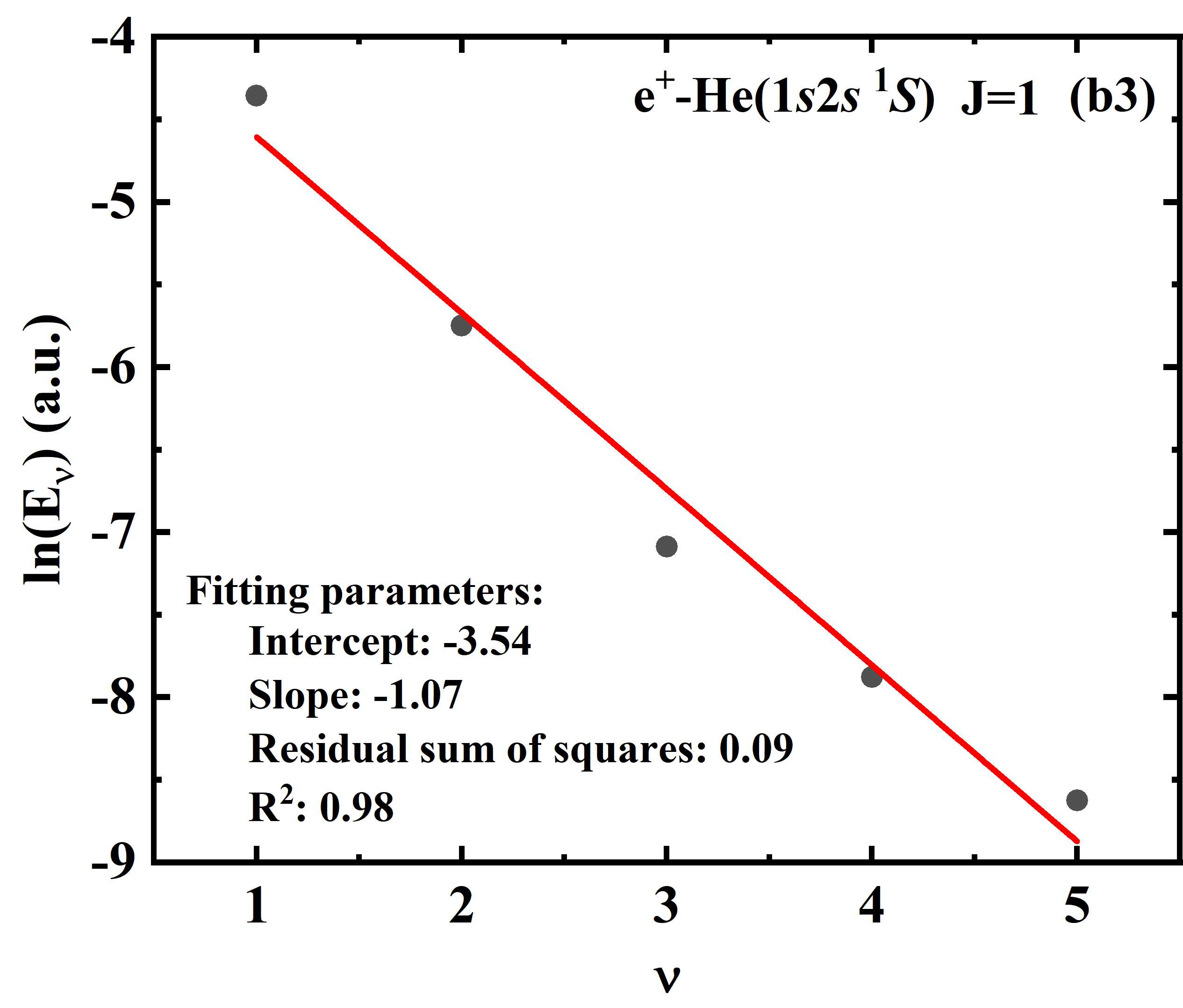}
		\label{fig8b3}
	}\\
	\renewcommand{\thesubfigure}{(c\arabic{subfigure})}
	\setcounter{subfigure}{0}
	\subfigure{
		\includegraphics[width=0.305\textwidth]{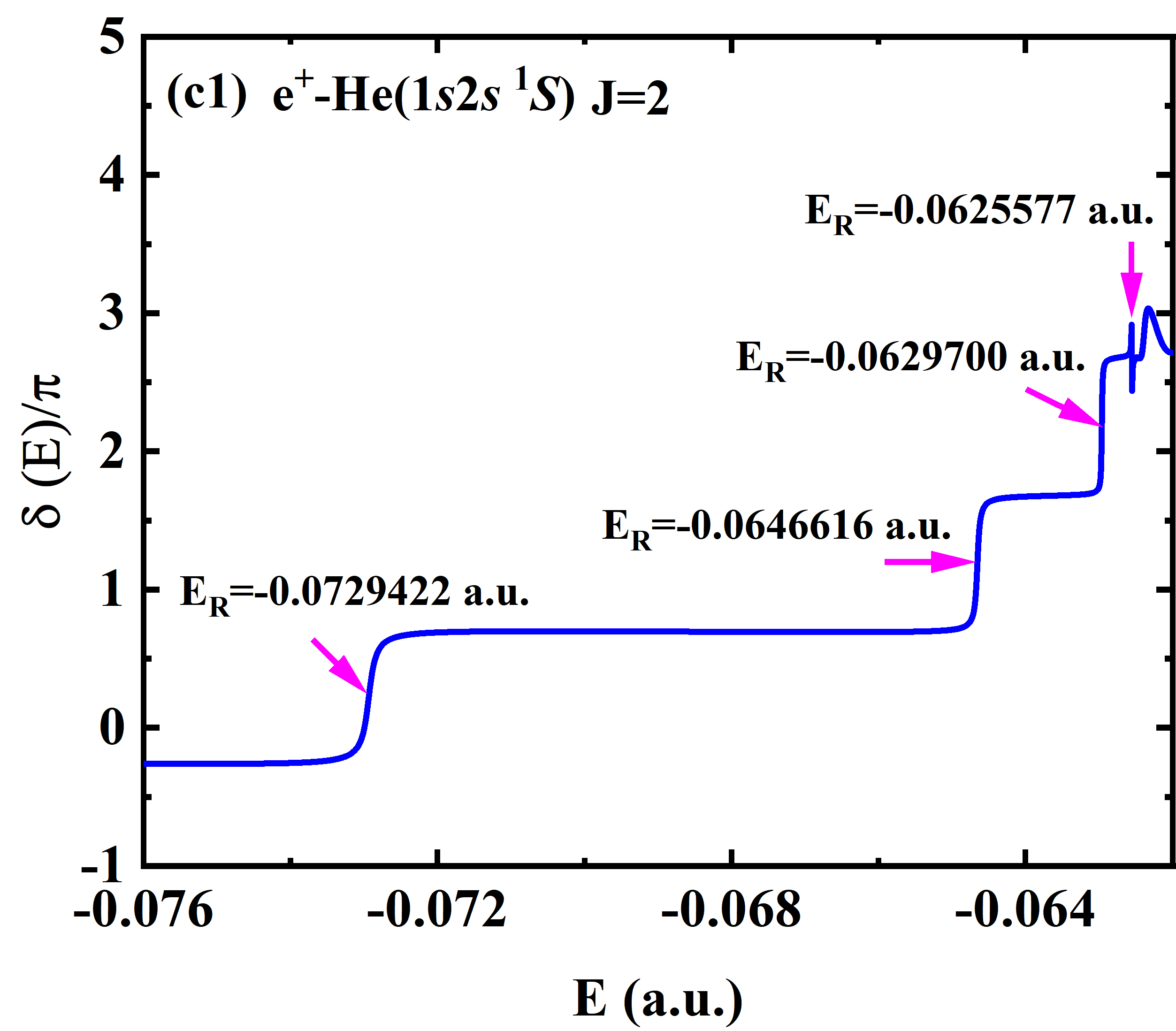}
		\label{fig8c1}
	}
	\subfigure{
		\includegraphics[width=0.32\textwidth]{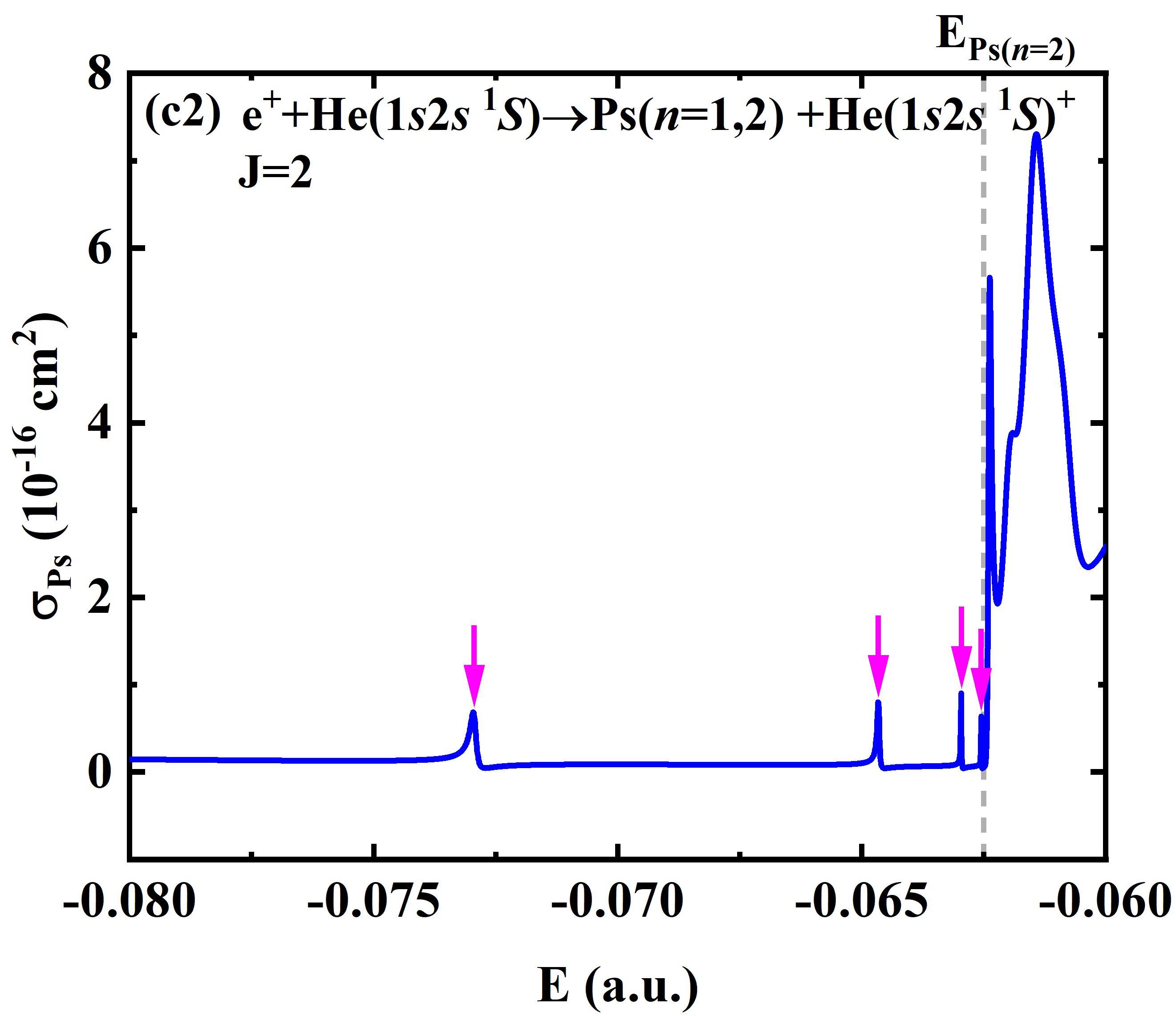}
		\label{fig8c2}
	}
	\subfigure{
		\includegraphics[width=0.315\textwidth]{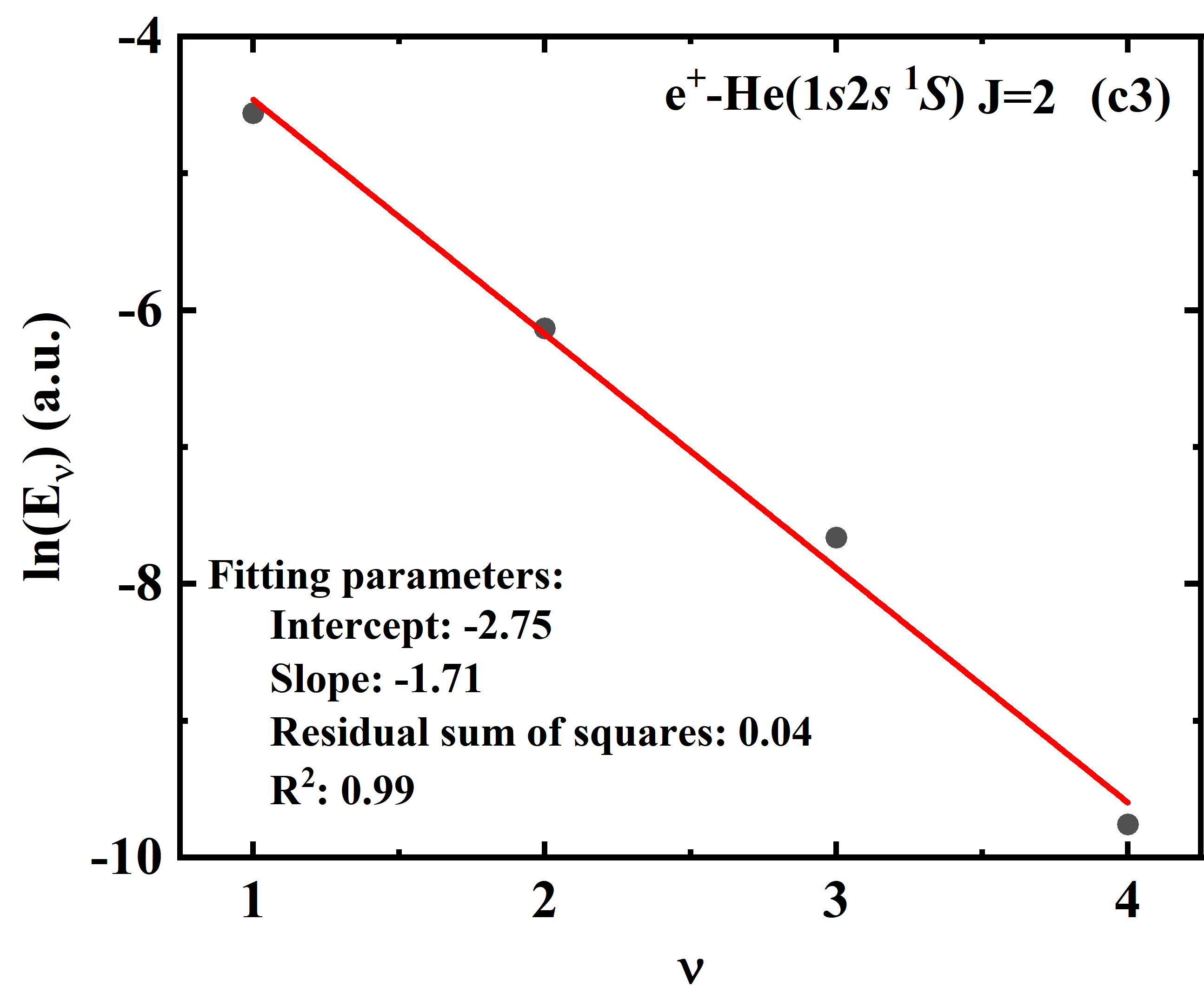}
		\label{fig8c3}
	}\\
	\renewcommand{\thesubfigure}{(d\arabic{subfigure})}
\setcounter{subfigure}{0}
\subfigure{
	\includegraphics[width=0.315\textwidth]{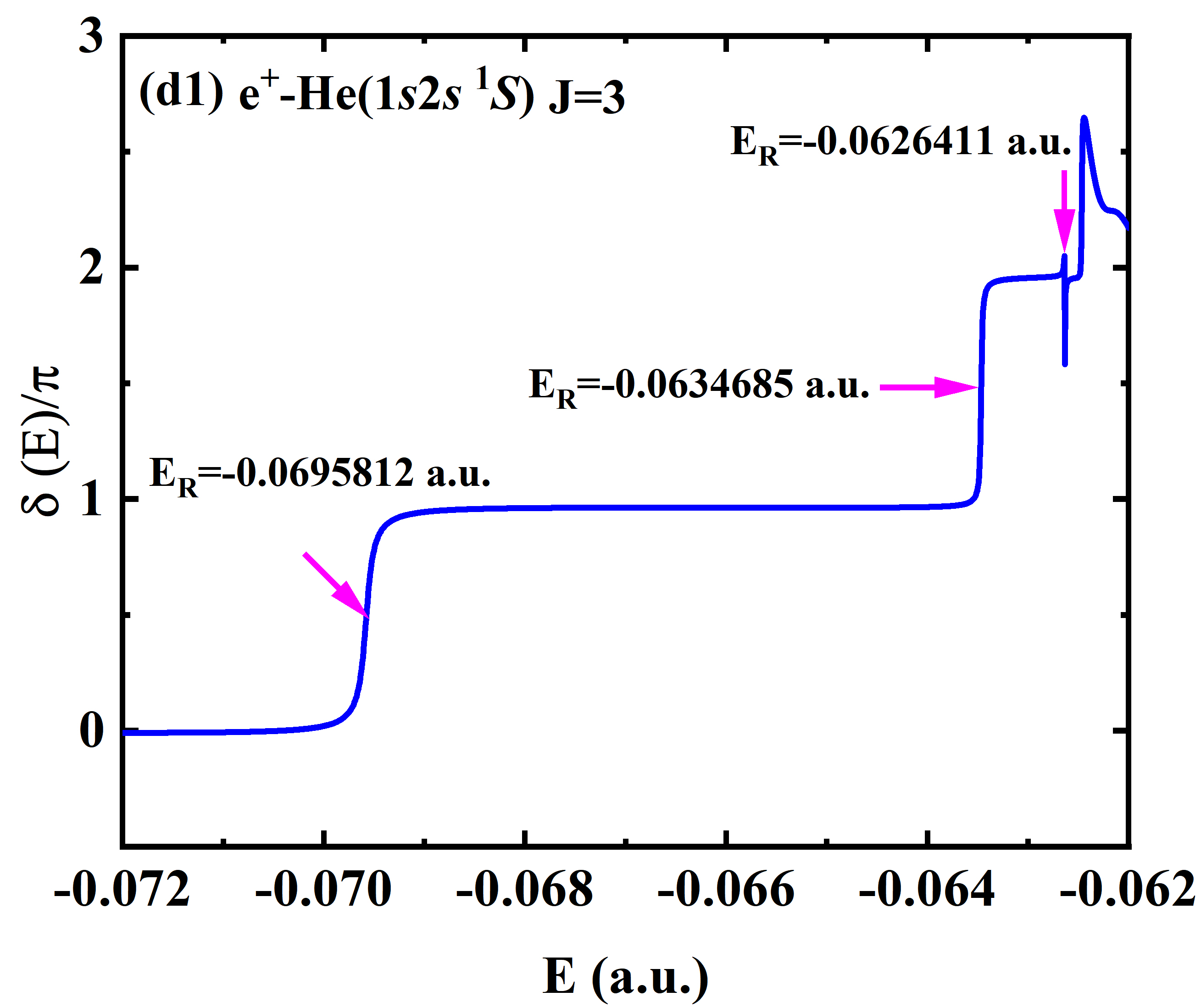}
	\label{fig8d1}
}
\subfigure{
	\includegraphics[width=0.32\textwidth]{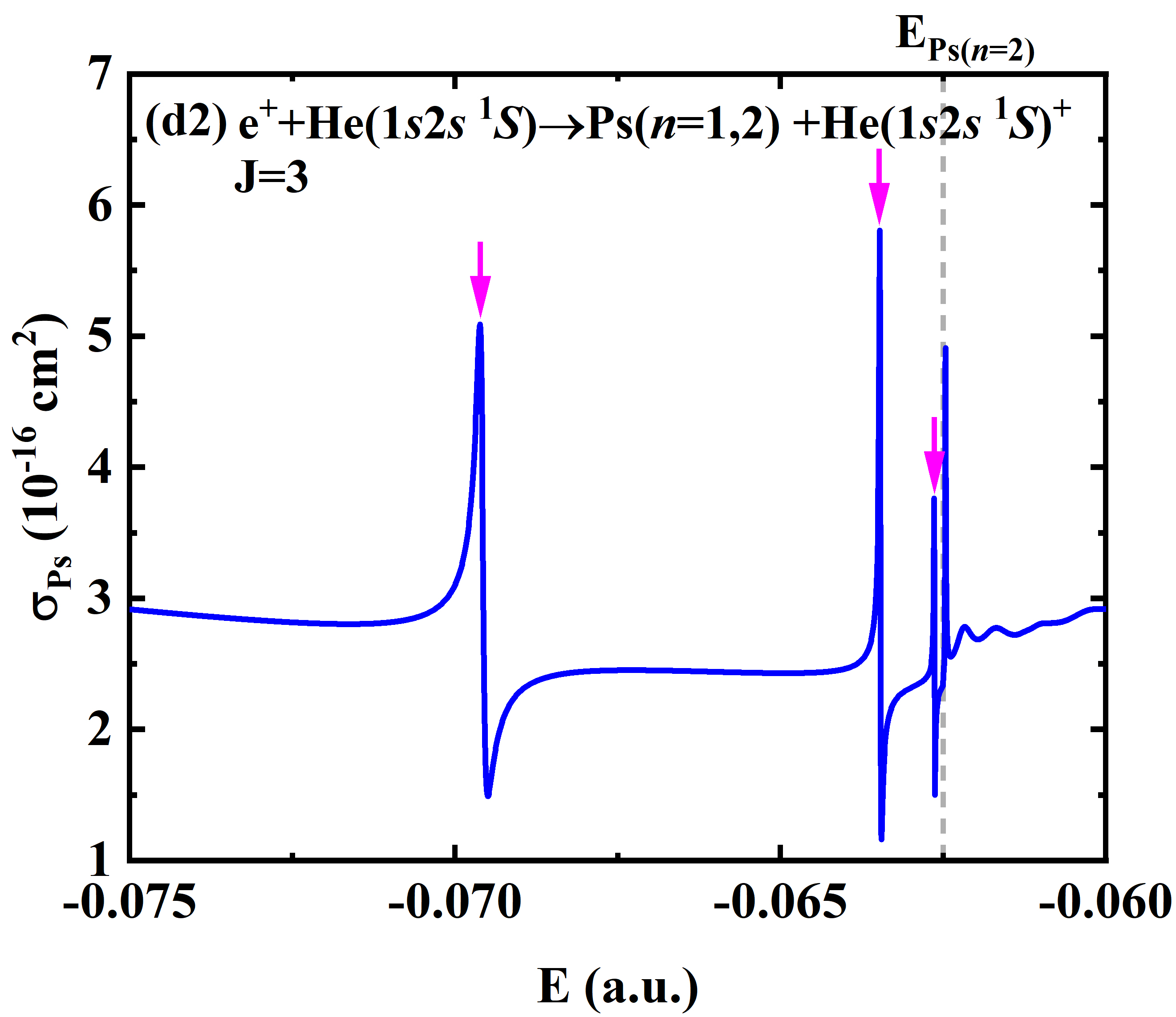}
	\label{fig8d2}
}
\subfigure{
	\includegraphics[width=0.315\textwidth]{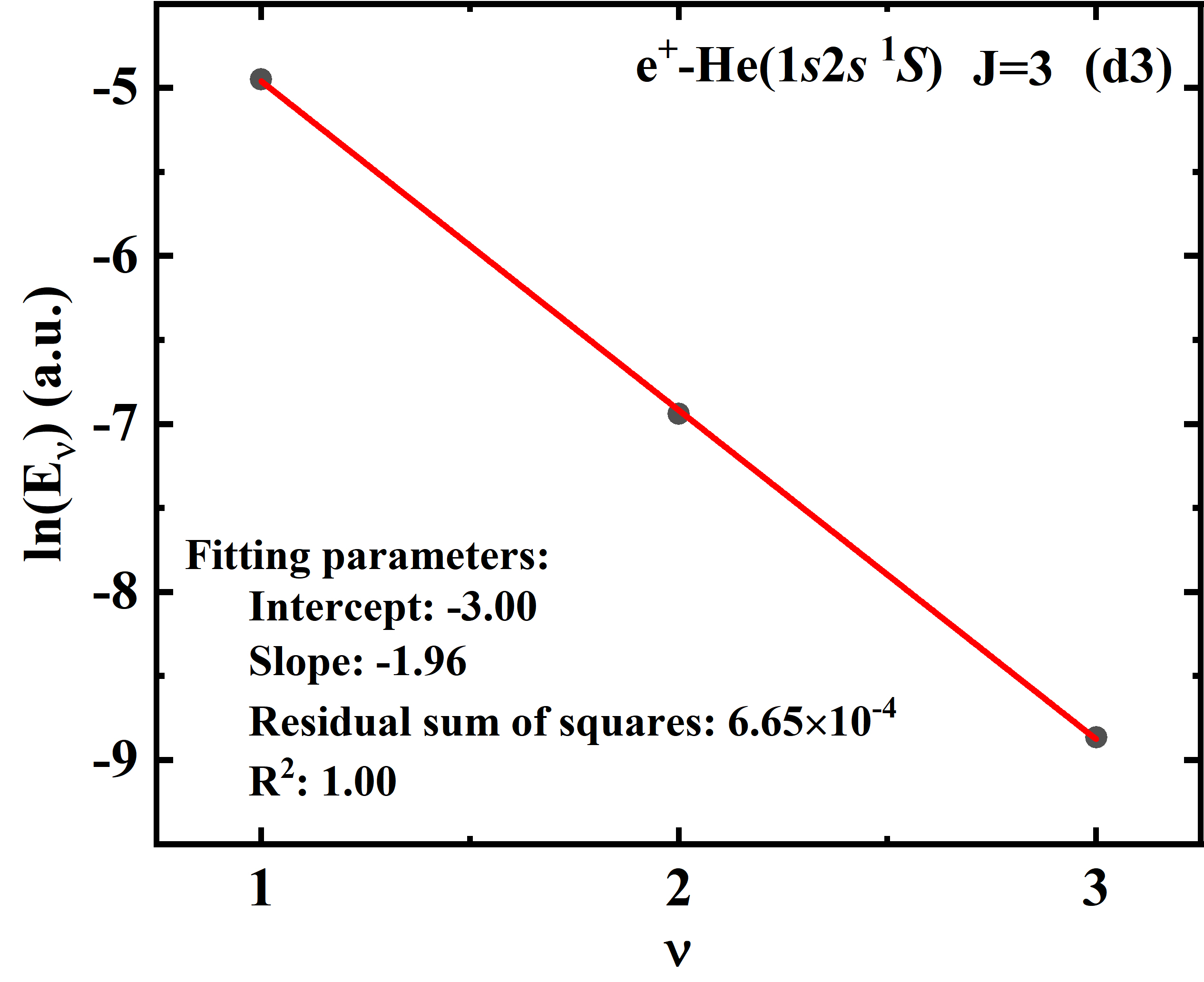}
	\label{fig8d3}
}
	\caption{(Color online) (a1), (b1), (c1), and (d1) Eigenphase sums as functions of energy; (a2), (b2), (c2), and (d2) Ps-formation cross sections for the e$^{\scriptscriptstyle+}$-He($1s2s\,^1S$) system with $J=0-3$ near the Ps($n=2$) threshold. The dashed lines indicate the Ps($n=2$) threshold and the arrows indicate the resonance positions. (a3), (b3), (c3), and (d3) Semilogarithmic plots of the resonance positions $E_{R}=E_\text{th}-E_{\nu}$ for Ps($n=2$)+He$^{\scriptscriptstyle+}(1s)$ below the Ps($n=2$) threshold. Straight lines represent the fits using Eq.~(\ref{lnEv}).}
\end{figure*}

Figures~\ref{fig7a1}--\ref{fig7c1} and Figs.~\ref{fig8a1}--\ref{fig8d1} present the eigenphase-sum spectra, Ps($n=1,2$) formation cross sections, and dipole resonance energies near the Ps($n=2$) threshold for the e$^{\scriptscriptstyle+}$-He($1s2s\,^3S$) ($J=0-2$) and e$^{\scriptscriptstyle+}$-He($1s2s\,^1S$) ($J=0-3$) systems, respectively. In both systems, the eigenphase sums exhibit rises of approximately $\pi$, indicating the presence of resonances, whose positions are marked by arrows. Corresponding resonance structures are also observed in the Ps($n=1,\,2$) formation cross sections (Figs.~\ref{fig7a2}--\ref{fig7c2} and Figs.~\ref{fig8a2}--\ref{fig8d2}). The extracted dipole resonance energies, $E_r$, are plotted on semi-logarithmic scales and fitted to Eq.~(\ref{lnEv}), as shown by the straight lines in Figs.~\ref{fig7a3}--\ref{fig7c3} and Figs.~\ref{fig8a3}--\ref{fig8d3}. Most resonance series consist of very narrow resonances, whose signatures in the scattering cross sections are weak and become discernible only upon magnification. The eigenphase-sum spectra further show that the resonance density increases rapidly as the energy approaches the Ps($n=2$) threshold. The identified resonances are listed in Table~\ref{t2}, and the ratios of successive resonance energies in Table~\ref{t3}.

\begin{table}[ht]
	\centering
	\caption{\label{t3}
		Ratios of successive resonance energies near the Ps($n=2$) threshold and the corresponding fitted scaling parameters $\alpha$ in the e$^{\scriptscriptstyle+}$--He($1s2s\,^{1,3}S$) systems for $J=0-3$.}
	\renewcommand{\arraystretch}{1.2}
	\setlength{\tabcolsep}{10pt}
	\begin{tabular}{ccccc}
		\hline\hline
		& \multicolumn{2}{c}{e$^{\scriptscriptstyle+}$-He($1s2s\,^3S$)}
		& \multicolumn{2}{c}{e$^{\scriptscriptstyle+}$-He($1s2s\,^1S$)} \\
		\cmidrule(lr){2-3} \cmidrule(lr){4-5}
		$\nu/\nu+1$
		& $E_{\nu}/E_{\nu+1}$
		& $\alpha$
        & $E_{\nu}/E_{\nu+1}$
        &$\alpha$\\
		\hline
		\multicolumn{5}{l}{\hspace{1em}$S$ wave} \\
		$1/2$ & 3.75 & 4.70& 3.68  & 7.83 \\
		$2/3$ & 3.86 && 3.61 &     \\
		$3/4$ & --   && 3.75 &       \\
		
		\multicolumn{5}{l}{\hspace{1em}$P$ wave} \\
		$1/2$ & 1.98& 6.74& 4.03  & 5.89 \\
		$2/3$ & 2.08&  & 3.82    &    \\
		$3/4$ & 4.24&  & 2.20    &    \\
		$4/5$ & --  & & 2.11     &   \\
		
		\multicolumn{5}{l}{\hspace{1em}$D$ wave} \\
		$1/2$ & 4.80 & 3.56& 4.83  & 3.67 \\
		$2/3$ & 7.08 & & 4.60   &    \\
		$3/4$ & --   & & 8.16 &        \\
		
		\multicolumn{5}{l}{\hspace{1em}$F$ wave} \\
		$1/2$ & 7.03& -- & 7.31    & 3.21 \\
		$2/3$ & --  &   & 6.86 &      \\
		
		\hline\hline
	\end{tabular}
\end{table}
\clearpage

%\bibliographystyle{apsrev_title.bst}
%\bibliography{Reference}

\end{document}